\documentclass{nsart_eng}
\usepackage{cite}
\usepackage{graphicx}
\usepackage{amsmath,amssymb}
\usepackage{multirow,longtable}

\usepackage{xcolor}

\year{2022} \volume{0} \nomer{0} \firstpage{1}

\let\leq\leqslant

\begin{document}

\title[CV-QKD security]
{Continuous-variable quantum key distribution: security analysis with trusted hardware noise against general attacks}

\author[R.~Goncharov, A.D.~Kiselev, E.~Samsonov, V.~Egorov]
{$^{1}$R.~Goncharov, $^{1}$A.D.~Kiselev, $^{1}$E.~Samsonov, $^{1}$V.~Egorov}

\address{
$^1$ ITMO University,\\
 Kronverkskiy, 49, St. Petersburg, 197101, Russia}

\email{rkgoncharov@itmo.ru}

\udk{530.145:535.12:681.7:53.082.5}


\begin{abstract}
In this paper, using the full security framework for continuous variable quantum key distribution (CV-QKD), we provide a composable security proof for the CV-QKD system in a realistic implementation. We take into account equipment losses and contributions from various components of excess noise and evaluate performance against collective and coherent attacks assuming trusted hardware noise. The calculation showed that the system remains operable at channel losses up to 10.2 dB in the presence of collective attacks and up to 7.5 dB in the presence of coherent ones.
\end{abstract}


\maketitle

\section{Introduction}

Quantum key distribution (QKD)~\cite{Pirandola2020} is a special method of generating a secure key between two parties, Alice and Bob, which will ensure the privacy of transmitted information in the era of the quantum computer. Historically, the first protocols to be presented were discrete variable (DV) ones~\cite{Bennett2014,Bennett1992}, where information was encoded in the state of a single photon: polarization, phase or time bin. However, over time, continuous variable (CV) protocols~\cite{Ralph1999,Cerf2001,Grosshans2003} have been introduced, which are considered more efficient, high-rate and cost-effective due to the use of homodyne/heterodyne detection systems instead of single photon detectors. 

Considering the security of QKD systems, one must take into account that each of them has a finite physical implementation that is not ideal, which opens up opportunities for the eavesdropper, Eve, to carry out a multiple attacks and extract part of the secret key. To prevent this threat, for each protocol, a complex system for assessing the information available to Eve and the acceptable level of errors is being developed. 

Currently, a fairly significant amount of work has been presented, covering the topic of security of CV-QKD protocols~\cite{Leverrier2010,Leverrier2015,Ghorai2019,Lin2019,Denys2021,Usenko2016,Laudenbach2017,Laudenbach2019}. Of the protocols most suitable for practical implementation, the GG02 protocol~\cite{Grosshans2003,Weedbrook2004a} stands out, for which the security is proven against coherent (general) attacks, taking into account the finite-key effects. Moreover, models of untrusted and trusted hardware noise are considered~\cite{Usenko2016}. The latter is preferable, since many security levels imply that Eve does not have access to Alice's and Bob's blocks, moreover, accounting for untrusted noise makes the protocol essentially unusable.

Thus, this paper will present a full security proof of CV-QKD on a realistic implementation with trusted hardware noise against
general attacks. In Section~\ref{sec:scheme} we describe an optical configuration of the CV-QKD scheme, in Sections~\ref{sec:protocol}--\ref{sec:gaussian} we give a description of the protocol in the trusted noise scenario and consider a possibility of specific attacks that go beyond general security proof framework. In Section~\ref{sec:parameters} we provide a technique of evaluation and monitoring of experimental parameters and in Section~\ref{sec:security} we clarify security analysis and estimate the finite-length secure key generation rate. In Section~\ref{sec:conclusion} we discuss the results and draw the appropriate conclusions.

\section{Optical CV-QKD scheme configuration}
\label{sec:scheme}
\begin{figure}[ht!]

\centering
\includegraphics[width=\textwidth]{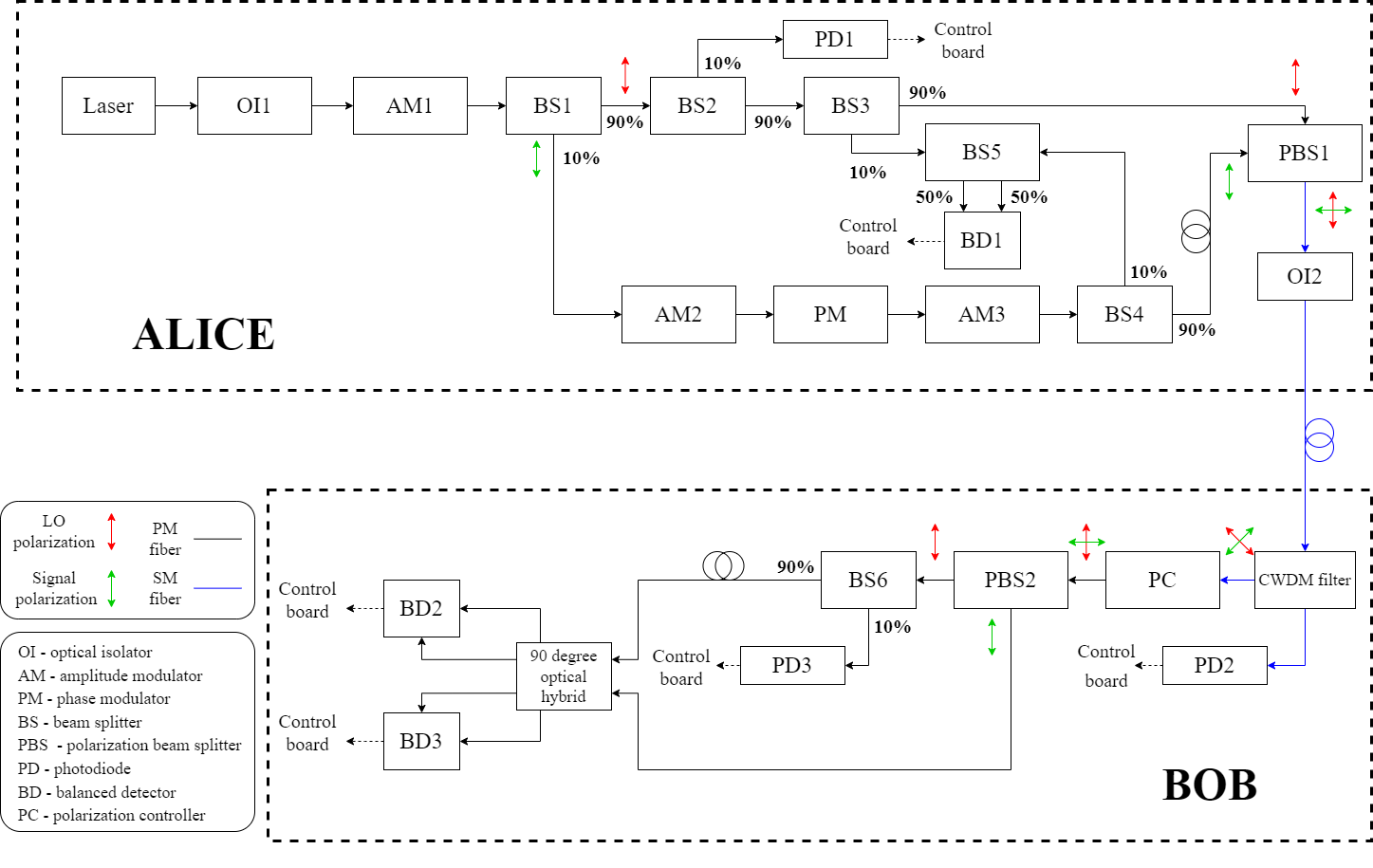}
\caption{Scheme of typical setup for CV-QKD with Gaussian modulation and heterodyne detection.}  
\label{fig:gen-scheme}
\end{figure}

The optical scheme of the described protocol is shown in Figure~\ref{fig:gen-scheme} and consists of the following blocks:
\begin{itemize}
    \item Alice block, in which the generation of signal states (Gaussian modulation) and a local oscillator (LO) is carried out, after which, by means of polarization-time multiplexing, they sent to Bob. Gaussian modulation itself can be done in two stages: amplitude modulation with a Rayleigh distribution and phase modulation with a uniform phase distribution; as a result, the the complex signal amplitude value will correspond to the Gaussian distribution~\cite{Huang2016}. It should also be noted that the need for reference signal, which is used in the phase protocols of the \mbox{DV-QKD~\cite{Bennett1992,Bennett1992a}}, is no longer necessary in this case. 
    \item Quantum channel block, which in this work is represented by an optical fiber. 
    \item Bob block, in which demultiplexing and heterodyne detection are carried out.
\end{itemize}

In the considered configuration Alice uses the same laser to generate LO and signal states modulated according to the Gaussian distribution, which then experience the procedure of polarization-time multiplexing before sending through the quantum channel. Bob then performs a demultiplexing procedure and heterodyne detection. Figure~\ref{fig:gen-scheme} also shows additional elements for monitoring.

In Alice block a continuous wave (CW) laser with a central wavelength of 1550~nm and a spectral line width of 100~kHz is used. The width of the spectral line makes it possible to estimate the coherence time, and also affects the amount of phase noise. Coherent detection systems require the use of sources with a narrow spectral line. The Bob block uses a polarization-maintaining optical fiber of the Panda type due to the sensitivity of the optoelectronic components to polarization and implementation polarization multiplexing. The optical isolator OI1 is used in order to avoid backlight due to reflection on the optical scheme elements in the laser module. An amplitude modulator AM1 is used to form identical optical pulses with a repetition rate of 50~MHz and a duration of 3~ns. Laser pulses then divided by amplitude into two arms using a 10/90 beam splitter BS1: the signal arm (10\%) and LO arm (90\%).

The second 90/10 beam splitter BS2 in LO arm is used to separate 10\% of the optical pulse power and direct it to the photodiode PD1, which allows to organize feedback with an amplitude modulator to control the power of LO.

In the signal arm, Gaussian modulation of optical pulses occurs using an amplitude modulator AM2 and a phase modulator PM. Amplitude modulation is realized according to the Rayleigh distribution with a given variance. Phase modulation is realized according to a uniform distribution in the range from 0 to 2$\pi$. The amplitude modulator AM3 is used as a fast attenuator with a wide range of extinction coefficient to set the average number of photons in an optical pulse. Every second pulse in the signal arm is a reference pulse and is not subjected to amplitude and phase modulation on amplitude modulators AM1 and AM2 and phase modulator PM. The modulators are controlled by a digital-to-analog converter (DAC), which is not indicated on Figure~\ref{fig:gen-scheme}.

In the signal and LO arms, 90:10 beam splitters BS3 and BS4 are used to separate part of the signal and LO pulses to control the power and variance (feedback for amplitude modulators AM1 and AM2) using a homodyne detector, which is carried out using 50/50 beam splitter BS5 and balanced detector BD1. For successful balanced detection, it is necessary that the optical paths of the signal and LO are identical.

In order to avoid the interaction of signal pulses and LO before their detection in Bob's side, time-division and polarization multiplexing is used. Time-division multiplexing is implemented by increasing length of the signal arm in Alice by an amount corresponding to half the pulse repetition period compared to LO arm. Polarization multiplexing is implemented using a polarization combiner PBS1 with a single-mode (SM) output, to which polarization maintaining fibers of two arms are connected with mutually orthogonal slow optical axes.

A SM optical isolator OI2 is used to prevent backlighting of Alice from the channel output. Alice and Bob are connected through a SM optical fiber of the G.652.D standard with a length of 25 km.

On Bob side the signal and LO enters the CWDM filter with a central wavelength of 1550~nm. The use of a CWDM filter is due to protection against backlight at a different wavelength from the channel input. The output of the CWDM filter, corresponding to the reflected light, goes to the monitor photodiode PD2, by the signal from which one can analyse the presence or absence of backlights. The output of the CWDM filter, corresponding to the transmitted light, is connected to the input of the polarization controller PC, which is used to compensate for polarization distortions in the SM optical fiber between the Alice and Bob. The input of a polarization beam splitter PBS2 is connected to the output of the polarization controller, which is used for polarization demultiplexing of the signal and LO. In LO arm there is a delay line for demultiplexing signals in time domain and also a 90:10 beam splitter BS6, which is used to organize feedback of the polarization controller: a part of LO pulse power (10\%) is sent to the photodiode PD3 to determine the optical power. The other output of the 90/10 beam splitter, which corresponds to 90\% optical power, is connected to LO input of the 90-degree optical hybrid. In the signal arm, after the output of the polarizing beam splitter PBS2, it is connected to the signal input of the 90-degree optical hybrid. The 90-degree optical hybrid has four outputs: two outputs correspond to the sum and difference amplitude of the signal field and LO with zero additional relative phase, and two outputs correspond to an additional phase equal to 90 degrees. The outputs of the 90-degree optical hybrid are connected to the inputs of two balanced detectors. The signal from balanced detectors is entered to analog-to-digital converter (ADC), which is not shown in the Figure~\ref{fig:gen-scheme}).

\section{GG02 protocol features}
\label{sec:protocol}
\subsection{Gaussian modulation of coherent states}
In the CV-QKD with Gaussian modulation~\cite{Weedbrook2004a,Grosshans2003}, Alice prepares coherent states (with a given value of amplitude and phase) with quadrature components $q$ and $p$ which are realizations of two independent and identically distributed random variables $\mathcal{Q}$ and $\mathcal{P}$, which have the same Gaussian distribution with zero mean and given variance
\begin{equation}
    \label{eq:dist-quad}
    \mathcal{Q} \sim \mathcal{P} \sim \mathcal{N}\left(0,\: \tilde{V}_{\mathrm{A}}\right),
\end{equation}
where $\tilde{V}_{\mathrm{A}}$ is a modulation variance.

Alice prepares a sequence of coherent states $|\alpha_{1}\rangle, \ldots|\alpha_{j}\rangle, \ldots,|\alpha_{N}\rangle$ of the form:
\begin{align}
|\alpha_{j}\rangle=|q_{j}+i p_{j}\rangle,\text{ for }
 q_j \in \mathcal{Q},\: p_j \in \mathcal{P}.
\end{align}

In this case, the equations for the eigenvalues are satisfied in shot noise units (SNU)
\begin{align}
\hat{a}|\alpha_{j}\rangle&=\alpha_{j}|\alpha_{j}\rangle,\\
\frac{1}{2}(\hat{q}+i \hat{p})|\alpha_{j}\rangle&=(q_{j}+i p_{j})|\alpha_{j}\rangle,
\end{align}
where $\hat{a}$ is a creation operator and $\hat{p}$, $\hat{q}$ are a quadrature operators.

The mean photon number in each individual state is estimated as follows
\begin{align}
\langle n_{j}\rangle=\langle\alpha_{j}|\hat{n}| \alpha_{j}\rangle=|\alpha_{j}|^{2}=q_{j}^{2}+p_{j}^{2}.
\end{align}
Given that $q_{j}$ and $p_{j}$ taken from the distribution in expression~\eqref{eq:dist-quad}, the mean photon number over the ensemble of states prepared by Alice is
\begin{equation}
\label{eq:mean-phot}
\langle n\rangle=\langle\mathcal{Q}^{2}\rangle+\langle\mathcal{P}^{2}\rangle=2 \tilde{V}_{\mathrm{A}}.
\end{equation}

To calculate the variance of the quadrature operator $V(\hat{q})=\langle\hat{q}^{2}\rangle-\langle\hat{q}\rangle^{2}$, it is necessary to find the averaged values:
\begin{align}
\langle\hat{q}\rangle &=\langle\alpha|\hat{q}| \alpha\rangle=0,\\
\notag
\langle\hat{q}^{2}\rangle &=\langle\alpha|\hat{q}^{2}| \alpha\rangle=\langle\alpha|(\hat{a}+\hat{a}^{\dagger})^{2}| \alpha\rangle =\\ \notag &=\langle\alpha|\hat{a}^{2}| \alpha\rangle+\langle\alpha|(\hat{a}^{\dagger})^{2}| \alpha\rangle+\langle\alpha|\hat{a} \hat{a}^{\dagger}| \alpha\rangle+\langle\alpha|\hat{a}^{\dagger} \hat{a}| \alpha\rangle =\\ \notag &=\alpha^{2}+(\alpha^{*})^{2}+\alpha^{*} \alpha+1+\alpha^{*} \alpha =\\ \notag &=q^{2}-p^{2}+2 i q p+q^{2}-p^{2}-2 i q p+2(q^{2}+p^{2})+1 =\\ &=4 q^{2}+1. \end{align}

Considering that the values $q$ and $p$ are realizations of the random variables $\mathcal{Q}$ and $\mathcal{P}$ respectively, we can write:
\begin{align}
    \label{eq:quad-variance}
    \langle\hat{q}^{2}\rangle&=4 \langle\mathcal{Q}^{2}\rangle+1=4\tilde{V}_{\mathrm{A}}+1,\\
    \langle\hat{p}^{2}\rangle&=4 \langle\mathcal{P}^{2}\rangle+1=4\tilde{V}_{\mathrm{A}}+1.
\end{align}
According to the equation~\eqref{eq:quad-variance}:
\begin{equation}
\label{eq:variance-full}
V\equiv V(\hat{q})=V(\hat{p})\equiv V=4 \tilde{V}_{\mathrm{A}}+1\equiv V_{\mathrm{A}}+1.
\end{equation}
In Eq.~\eqref{eq:variance-full} the transition from the variance of random variable corresponding to the quadrature distribution to the variance of quadrature operator is carried out. It should be noted that there is also a shot noise component in SNU (equal to one).

Combining~\eqref{eq:mean-phot} and~\eqref{eq:variance-full}, we can express the average number of photons over the ensemble in terms of the variance of quadrature operator
\begin{equation}
\label{eq:mean-phot-fin}
\langle n\rangle=\frac{1}{2}(V-1)=\frac{1}{2} V_{\mathrm{A}}.
\end{equation}

After the preparation stage, Alice sends the $|\alpha_{j}\rangle$ state to the Gaussian quantum channel, after which Bob performs coherent detection and decodes information about the sent state in the case of heterodyne detection or the projection of its quadrature components in the case of homodyne detection. It should be emphasized that in this paper only heterodyne detection is considered.

\subsection{Gaussian sequence processing}
This subsection briefly describes the stages of classical data post-processing, which means Alice's modulation and Bob's detection results data.

The first step is sifting. Despite sifting is not implied in the case of heterodyne detection (in which the sequence after the distribution session and before error correction can be considered as a sifted key), it is worth mentioning that in CV-QKD with homodyne detection, Alice and Bob choose the bases they use to prepare and measure the states respectively, using independently and identically distributed generated random bits. In these cases, the sifting step eliminates all uncorrelated signals when different bases were used for preparation and measurement. The presence of signal bases is more typical for CV-QKD protocols with discrete modulation, however, in protocols with Gaussian modulation and homodyne detection, sifting means discarding the quadrature not measured by Bob.

The second step is parameter estimation. After transmitting and detecting a sequence of states, legitimate parties compare a random subset of their data. This comparison allows estimating the quantum channel parameters: transmittance and excess noise of the channel, from which they can calculate a mutual information $I_{\mathrm{AB}}$ and evaluate an information $\chi$ available to Eve. If $\chi$ is greater than $\beta I_{\mathrm{AB}}$, where $\beta \in [0,\:1]$ is a reconciliation efficiency, the protocol is aborted at this point.

If $\beta I_{\mathrm{AB}}>\chi$, users go to the third, information reconciliation step, which is a form of error correction procedure.

The fourth step is confirmation. After the reconciliation procedure, legitimate parties perform a confirmation step using a family of universal hash functions~\cite{Carter1979} to limit the chance that error correction fails: Alice or Bob chooses one particular hash function with uniform probability and announces its choice over the classical channel. Users apply this hash function to their key to get a hash code. Subsequently, Alice and Bob exchange and compare their hash codes. If the values are different, the keys are considered compromised and the protocol is aborted; if the values are equal, then it is considered that an upper bound on the probability that the keys are not identical has been obtained. This error rate depends on the length of the hash codes and the type of hash functions used.

The fifth and final step is privacy amplification. After successfully passing the confirmation stage, Alice and Bob will have the same bit string with a very high probability. However, Eve has some information about the key, so to reduce the chance that she successfully guesses part of the key to an acceptable value, users perform a privacy amplification protocol by applying a seeded randomness extraction algorithm to their bit strings, which uses a family of 2-universal hash functions.

\subsection{Quantum channel description}
The Gaussian quantum channel is characterized by the transmittance coefficient (taking into account losses directly in the channel, losses in the equipment and detection efficiency) and noise (on Alice side, in the channel, and on Bob side). The potential advantage of Eve depends on both characteristics. The noise in the channel can be expressed as~\cite{Pirandola2020,Scarani2009,Laudenbach2017}
\begin{equation}
\label{eq:channel-n}
\Xi_{\mathrm{ch}}= \frac{1-T_{\mathrm{ch}}}{T_{\mathrm{ch}}}+\xi_\mathrm{{A}},
\end{equation}
where $T_{\mathrm{ch}}$ is a quantum channel transmittance, $\xi_{\mathrm{A}}$ is a excess noise (in SNU).

The excess noise itself includes variances of all noise sources
\begin{equation}
\xi_{\mathrm{A}}=\xi_{\mathrm {modul,\:A}}+\xi_{\mathrm {Raman,\:A}}+\xi_{\mathrm {phase,\:A}}+\cdots,
\end{equation}
where $\xi_{\mathrm{modul,\:A}}$ is a modulation noise, $\xi_{\mathrm{Raman,\:A}}$ is a Raman noise, and $\xi_{\mathrm{phase,\:A}}$ is a phase noise. 

Similarly, the detector noise can be estimated as~\cite{Scarani2009,Pirandola2020,Laudenbach2017,Laudenbach2019}
\begin{align}
\Xi_{\mathrm{det}}=\frac{1-T_{\mathrm{rec}}\eta_{\mathrm{det}}}{T_{\mathrm{rec}}\eta_{\mathrm{det}}}+\frac{v_{\mathrm{el}}}{T_{\mathrm{rec}}\eta_{\mathrm{det}}}
\end{align}
where $\eta_{\mathrm{det}}$ is a balanced detector efficiency, $v_{\mathrm{el}}$ is an electronic noise of balanced detector and $T_{\mathrm{rec}}$ is a transmittance coefficient responsible for losses in the receiver module. 

For a more convenient notation, it can be written as
\begin{align}
    T_{\mathrm{det}} \equiv T_{\mathrm{rec}}\eta_{\mathrm{det}}.
\end{align}

The total noise related to the channel input is then determined by the sum of the channel noise and the detector noise normalized to $T_{\mathrm{ch}}$
\begin{equation}
\Xi=\Xi_{\mathrm{ch}}+\frac{1}{T_{\mathrm{ch}}} \Xi_{\mathrm{det}}.
\end{equation}

After the signal state passing through a channel with noise and losses, Bob measures the total variance of the quadrature operator as~\cite{Pirandola2020,Scarani2009,Laudenbach2017,Laudenbach2019}
\begin{align}
    \notag
    V_{\mathrm{B}}&= V\left(\hat{q}_{\mathrm{B}}\right)=V\left(\hat{p}_{\mathrm{B}}\right)=T_{\mathrm{ch}}T_{\mathrm{det}}(V+\Xi)=\\\notag
    &=T_{\mathrm{ch}}T_{\mathrm{det}}\left(V+\frac{1-T_{\mathrm{ch}}}{T_{\mathrm{ch}}}+\xi_{\mathrm{A}}+\frac{1}{T_{\mathrm{ch}}}\left(\frac{1-T_{\mathrm{det}}}{T_{\mathrm{det}}}+\frac{v_{\mathrm{el}}}{T_{\mathrm{det}}}\right)\right) =\\\notag &=T_{\mathrm{ch}}T_{\mathrm{det}} V-T_{\mathrm{ch}} T_{\mathrm{det}}+T_{\mathrm{ch}} T_{\mathrm{rec}}\eta_{\operatorname{det}} \xi_{\mathrm{A}}+1+v_{\mathrm{el}} \equiv\\ \notag &\equiv T V-T+T \xi_{\mathrm{A}}+1+v_{\mathrm{el}} =\\ &=T(V-1)+T \xi_{\mathrm{A}}+1+v_{\mathrm{el}}.
\end{align}

Since the parameter $v_{\mathrm{el}}$ is the noise variance in SNU and can be considered stochastically independent of other noise sources, it can be considered as another component of excess noise, i.e. $v_{\mathrm{el}}\equiv \xi_{\mathrm{det}}$, thus
\begin{align}
\label{eq:xi-tot}
T \xi_{\mathrm{A}}+\xi_{\mathrm{det}}&=T\left(\xi_{\mathrm{A}}+\frac{1}{T} \xi_{\mathrm{det}}\right)\equiv T \xi_{\mathrm{tot,\:A}}= \xi_{\mathrm{tot,\:B}}\equiv \xi,\\
\label{eq:vb}
V_{\mathrm{B}} &=T(V-1)+1+\xi=T V_{\mathrm{A} }+1+\xi.
\end{align}

\subsection{Signal-to-noise ratio and mutual information}
The signal-to-noise ratio is expressed as
\begin{equation}
    \label{eq:snr}
    \mathrm{SNR}=\frac{P_{\mathrm{S}}}{P_{\mathrm{N}}},
\end{equation}
where $P_{\mathrm{S}}$ is a total signal power and $P_{\mathrm{N}}$ is a total noise power.

The purposed model makes it possible to separate the signal and noise components in the observed by Bob variance of the quadrature operator
\begin{equation}
\label{eq:var-bob}
V_{\mathrm{B}}=\frac{T}{\mu} V_{\mathrm{A}}+1+\frac{\xi}{\mu},
\end{equation}
where $\mu \in \{1;\:2\}$ is a homodyne/heterodyne detection system parameter respectively.

Thus the signal-to-noise ratio for the purposed protocol is
\begin{equation}
\label{eq:snr-2}
\mathrm{SNR}=\frac{\frac{1}{\mu} T V_{\mathrm{A}}}{1+\frac{1}{\mu} \xi}.
\end{equation}

Mutual information between Alice and Bob in this case is evaluated as~\cite{Laudenbach2017}
\begin{equation}
\label{eq:mutual-inf}
I_{\mathrm{\mathrm{AB}}}=\frac{\mu}{2} \log _{2}(1+\mathrm{SNR})=\frac{\mu}{2} \log _{2}\left(1+\frac{\frac{1}{\mu} T V_{\mathrm{A}}}{1+\frac{1}{\mu} \xi}\right).
\end{equation}

As it was already mentioned, the purposed protocol assumes a heterodyne detection method, i.e. $\mu=2$. According to the equation~\eqref{eq:mutual-inf}, despite the increase in mutual information by a factor of two (two quadratures per message are detected at once, instead of one), the signal-to-noise ratio decreases. Obviously (estimating the rate of increase of logarithmic functions), the advantage of heterodyne detection in terms of estimating mutual information will be observed only at large~$TV_{\mathrm{A}}$.

\section{Trusted hardware noise. Holevo bound}
\label{sec:trusted}
After variable $\rho_{\mathrm{AB}}$ has been removed from the shared state equation, it can be viewed as a pure two-particle state with a common Alice and Bob on one side and Eve on the other. As such, it can be written in terms of the Schmidt decomposition
\begin{equation}
\left|\Psi_{\mathrm{ABE}}\right\rangle=\sum_{i} \sqrt{\lambda_{i}}\left|\psi_{i}\right\rangle_{\mathrm{AB}}\left|\phi_{i}\right\rangle_{\mathrm{E}},
\end{equation}
where $\lambda_{j}$ is a real non-negative number.

Taking a partial trace over subsystems gives
\begin{align}
\operatorname{Tr}_{\mathrm{E}}\rho_{\mathrm{AB}}&=\rho_{\mathrm{AB}}=\sum_{i} \lambda_{i}|\psi_{i}\rangle_{\mathrm{AB}}\langle \psi_{i}|, \\ \operatorname{Tr}_{\mathrm{AB}}\rho_{\mathrm{AB}}&=\rho_{\mathrm{E}}=\sum_{i} \lambda_{i}|\phi_{i}\rangle_{\mathrm{E}}\langle \phi_{i}|.
\end{align}

The von Neumann entropy depends only on the $\lambda_{i}$ components, which, due to the Schmidt decomposition, are the same for $\rho_{\mathrm{AB}}$ and $\rho_{\mathrm{E}}$. Therefore, the von Neumann entropy of Eve is the same as the entropy shared by Alice and Bob
\begin{equation}
S_{\mathrm{E}}=S_{\mathrm{AB}}=-\sum_{i} \lambda_{i} \log_{2} \lambda_{i}.
\end{equation}

Then the following transformation is obvious
\begin{equation}
\label{eq:chi-final}
\chi_{\mathrm{E}} =S_{\mathrm{E}}-S_{\mathrm{E} \mid \mathrm{B}}=S_{\mathrm{AB}}-S_{\mathrm{A} \mid \mathrm{B}}.
\end{equation}

In the trusted noise model~\cite{Usenko2016}, the noise coming from the Alice's and Bob's equipment is assumed to be trusted, that is, Eve cannot manipulate it. The same applies to equipment losses. In this context, it is necessary to clarify the equation~\eqref{eq:xi-tot}
\begin{equation}
    \label{eq:excess-noise-full}
    \xi=T\xi_{\mathrm{pr}}+T_{\mathrm{det}}\xi_{\mathrm{ch}}+\xi_{\mathrm{rec}},
\end{equation}
where $\xi_{\mathrm{pr}}$ is an Alice excess noise, $\xi_{\mathrm{ch}}$ is a channel excess noise, and $\xi_{\mathrm{rec}}$ is a Bob excess noise.

Assuming that the detection devices are well calibrated and reliable, $T_{\mathrm{rec}}$ and $\xi_{\mathrm{rec}}$ are beyond Eve's influence. Then the covariance matrix describing its von Neumann entropy prior to measurement by Bob is
\begin{equation}
\label{eq:cov-trusted}
\Sigma_{\mathrm{AB}}^{\text {trusted rec. }}=\left(\begin{array}{cc}V \mathbf{1}_{2} & \sqrt{T_{\mathrm{ch}}(V^{2}-1)} \sigma_{z} \\ \sqrt{T_{\mathrm{ch}}(V^{2}-1)} \sigma_{z} & \left(T_{\mathrm{ch}}(V-1)+1+\xi_{\mathrm{ch}}\right) \mathbf{1}_{2}\end{array}\right).
\end{equation}

The matrix itself can be represented in the form
\begin{equation}
\label{eq:matrix-form}
\left(\begin{array}{ll}a \mathbf{1}_{2} & c \sigma_{z} \\ c \sigma_{z} & b \mathbf{1}_{2}\end{array}\right).
\end{equation}

The symplectic eigenvalues of this matrix are expressed as
\begin{align}
\label{eq:eigen-formula}
v_{1,\:2}=\frac{1}{2}(z \pm(b-a)),
\end{align}
where $z=\sqrt{(a+b)^{2}-4 c^{2}}$.

Although in the trusted noise model the quantities $T_{\mathrm{rec}}$ and $\xi_{\mathrm{rec}}$ do not contribute to $S_{\mathrm{E}}$, though they do affect Alice measurements and, consequently, Eve entropy $S_{\mathrm{E} \mid \mathrm{B}}$.

In the trusted noise scenario, the eavesdropper can only manipulate the state in the channel and carry out purifying in the same place. This means that the state of the system must be viewed through three distinct subsystems in the entanglement base scenario. Let the state "Alice-Bob-Eve" consist of two entangled states and a thermal state, each of which is uniquely determined by its variance: one entangled state $\mathrm{EPR}_{\mathrm{AB}}$ with variance $V$ used for key exchange between Alice and Bob, one entangled state $\mathrm{EPR}_{\mathrm{ch}}$ with variance $W_{\mathrm{ch}}$ for modeling noise and loss in the quantum channel and thermal state $\mathrm{Th}_{\mathrm{rec}}$ with variance $W_{\mathrm{rec}}$ to simulate the noise and losses of the receiver. The beam splitters, one with $T_{\mathrm{ch}}$ transmittance and one with $T_{\mathrm{rec}}$ transmittance, mix the initial Bob's entangled state modes with the channel state and thermal state modes, respectively. The general state before the action of beam splitters can be represented by the covariance matrix
\begin{align}
\notag \Sigma_{\mathrm{tot},\:0}&=\mathrm{EPR}_{\mathrm{AB}} \oplus \mathrm{EPR}_{\mathrm{ch}} \oplus \mathrm{Th}_{\mathrm{rec}}=\\&=\left(\begin{array}{ccccc}V \mathbf{1}_{2} & \sqrt{V^{2}-1} \sigma_{z} & 0 & 0 & 0 \\ \sqrt{V^{2}-1} \sigma_{z} & V \mathbf{1}_{2} & 0 & 0 & 0 \\ 0 & 0 & W_{\mathrm{ch}} \mathbf{1}_{2} & \sqrt{W_{\mathrm{ch}}^{2}-1} \sigma_{z} & 0 \\ 0 & 0 & \sqrt{W_{\mathrm{ch}}^{2}-1} \sigma_{z} & W_{\mathrm{ch}} \mathbf{1}_{2} & 0 \\ 0 & 0 & 0 & 0 & W_{\mathrm{rec}} \mathbf{1}_{2}\end{array}\right).
\end{align}

It can be noted here that the eavesdropper attack on the quantum channel involves only two entangled states, $\mathrm{EPR}_{\mathrm{AB}}$ and $\mathrm{EPR}_{\mathrm{ch}}$, which guarantees purifying as part of an attack. Unitary equivalence with complete purifying thus makes it easier to express the state on Bob's side: in the general case, it must be represented by another entangled state~\cite{Laudenbach2019}.

The beam splitter in the channel affects Alice's mode and one of the $\mathrm{EPR}_{\mathrm{ch}}$ state modes; the second beam splitter affects Bob's mode and the thermal state simulating the detection module
\begin{align}\mathrm{BS}_{\mathrm{ch}}&=\left(\begin{array}{ccccc}\mathbf{1}_{2} & 0 & 0 & 0 & 0 \\ 0 & \sqrt{T_{\mathrm{ch}}} \mathbf{1}_{2} & \sqrt{1-T_{\mathrm{ch}}} \mathbf{1}_{2} & 0 & 0 \\ 0 & -\sqrt{1-T_{\mathrm{ch}}} \mathbf{1}_{2} & \sqrt{T_{\mathrm{ch}}} \mathbf{1}_{2} & 0 & 0 \\ 0 & 0 & 0 & \mathbf{1}_{2} & 0 \\ 0 & 0 & 0 & 0 & \mathbf{1}_{2}\end{array}\right), \\ \mathrm{BS}_{\mathrm{rec}}&=\left(\begin{array}{ccccc}\mathbf{1}_{2} & 0 & 0 & 0 & 0 \\ 0 & \sqrt{T_{\mathrm{det}}} \mathbf{1}_{2} & 0 & 0 & \sqrt{1-T_{\mathrm{det}}} \mathbf{1}_{2} \\ 0 & 0 & \mathbf{1}_{2} & 0 & 0 \\ 0 & 0 & 0 & \mathbf{1}_{2} & 0 \\ 0 & -\sqrt{1-T_{\mathrm{det}}} \mathbf{1}_{2} & 0 & 0 & \sqrt{T_{\mathrm{det}}} \mathbf{1}_{2}\end{array}\right).
\end{align}
Denoting the sequence of both beam splitters as $\mathrm{BS}_{\mathrm{tot}}=\mathrm{BS}_{\mathrm{rec}}\mathrm{BS}_{\mathrm{ch}}$, the total quantum state is transformed as follows
\begin{equation}
\label{eq:symp-bs}
\Sigma_{\mathrm {tot }}=\mathrm{B S}_{\mathrm {tot }} \Sigma_{\mathrm {tot }, 0} \mathrm{B S}_{\mathrm {tot }}^{T}.
\end{equation}

To simplify expressions for the covariance matrix, it should be considered block by block. Thus, the block describing Alice-Bob subsystem is transformed as follows
\begin{equation}
\Sigma_{\mathrm{AB}}=\left(\begin{array}{cc}V \mathbf{1}_{2} & \sqrt{T_{\mathrm{ch}}} \sqrt{T_{\mathrm{det}}} \sqrt{V^{2}-1} \sigma_{z} \\ \sqrt{T_{\mathrm{ch}}} \sqrt{T_{\mathrm{det}}} \sqrt{V^{2}-1} \sigma_{z} & \left(\begin{array}{c}T_{\mathrm{ch}} T_{\mathrm{det}} V \\ +\left(1-T_{\mathrm{ch}}\right) T_{\mathrm{det}} W_{\mathrm{ch}} \\ +\left(1-T_{\mathrm{det}}\right) W_{\mathrm{rec}}\end{array}\right)\end{array}\right).
\end{equation}
Let the variances for entangled states be defined as
\begin{align}
W_{\mathrm{ch}}&=\frac{\xi_{\mathrm{ch}}}{1-T_{\mathrm{ch}}}+1, \\ W_{\mathrm{rec}}&=\frac{\xi_{\mathrm{rec}}}{1-T_{\mathrm{det}}}+1.
\end{align}

Then the variance of Bob's quadrature operator is
\begin{align}\notag V_{\mathrm{B}} &=T_{\mathrm{ch}} T_{\mathrm{det}}(V-1)+1+T_{\mathrm{det}} \xi_{\mathrm{ch}}+\xi_{\mathrm{rec}}= \\ &=T_{\mathrm{ch}} T_{\mathrm{det}}(V-1)+1+\xi_{\mathrm{ch}, B}+\xi_{\mathrm{rec}}. \end{align}

Final expression for Alice-Bob block
\begin{equation}
\Sigma_{\mathrm{A B}}=\left(\begin{array}{cc}V \mathbf{1}_{2} & \sqrt{T} \sqrt{V^{2}-1} \sigma_{z} \\ \sqrt{T} \sqrt{V^{2}-1} \sigma_{z} & \left(T(V-1)+1+\xi\right) \mathbf{1}_{2}\end{array}\right).
\end{equation}

Eve block is described by the matrix
\begin{equation}
\label{eq:cov-mat-e-1}
\Sigma_{\mathrm{E}}=\left(\begin{array}{cc}\left(\left(1-T_{\mathrm{ch}}\right) V+T_{\mathrm{ch}} W_{\mathrm{ch}}\right) \mathbf{1}_{2} & \sqrt{T_{\mathrm{ch}}} \sqrt{\mathrm{W}_{\mathrm{ch}}^{2}-1} \sigma_{z} \\ \sqrt{T_{\mathrm{ch}}} \sqrt{W_{\mathrm{ch}}^{2}-1} \sigma_{z} & W_{\mathrm{ch}} \mathbf{1}_{2}\end{array}\right).
\end{equation}

This matrix can be described in the form given by Eq.~\eqref{eq:matrix-form}, so its symplectic eigenvalues can be calculated by Eq.~\eqref{eq:eigen-formula}. It can be verified that the entropy of Eve $S_{\mathrm{E}}$ is the same as the entropy shared by Alice and Bob and obtained from their mutual covariance matrix in the trusted receiver noise scenario from Eq.~\eqref{eq:cov-trusted}, which is expected when Eve purifies the state of Alice and Bob, i.e.
\begin{equation}
S_{\mathrm{E}}\equiv S\left(\Sigma_{\mathrm{E}}\right)=S\left(\Sigma_{\mathrm{AB}}^{\text {trusted rec. }}\right)\equiv S_{\mathrm{AB}}.
\end{equation}
Now, in order to get $S_{\mathrm{E\mid B}}$, we have to calculate $\Sigma_{\mathrm{tot \mid B}}$, i.e. the covariance matrix of the common state of the remaining modes after the projective measurement of the receiver mode. It is convenient to represent $\Sigma_{\mathrm{tot}}$ so that the Alice's mode is located in the last row and column. This can be done by using a permutation matrix, which allows to rearrange the third and fourth rows (columns) down (to the right) when multiplied by $\Sigma_{\mathrm{tot}}$ from the left (right)~\cite{Laudenbach2019}:
\begin{align}
P_{3,\:4 \rightarrow 9,\:10}&=\left(\begin{array}{ccccc}\mathbf{1}_{2} & 0 & 0 & 0 & 0 \\ 0 & 0 & \mathbf{1}_{2} & 0 & 0 \\ 0 & 0 & 0 & \mathbf{1}_{2} & 0 \\ 0 & 0 & 0 & 0 & \mathbf{1}_{2} \\ 0 & \mathbf{1}_{2} & 0 & 0 & 0\end{array}\right), \\
\Sigma_{\mathrm{tot}}^{\prime}&=P_{3,4 \rightarrow 9,10} \Sigma_{\mathrm{tot}} P_{3,4 \rightarrow 9,10}^{T}.
\end{align}

Since $P_{3,\:4 \rightarrow 9,\:10} P_{3,\:4 \rightarrow 9,\:10}^{T}=\mathbf{1}$, the above permutation is a similarity transformation and therefore leaves the eigenvalues of the matrix $\Sigma_\mathrm{tot}$ invariant. The covariance matrix itself now looks like this:
\begin{equation}
\Sigma_{\mathrm{tot}}^{\prime}=\left(\begin{array}{cc}\Sigma_{\mathrm{A},\:\mathrm{ch},\:\mathrm{rec}} & \Sigma_{\mathrm{C}} \\ \Sigma_{\mathrm{C}}^{T} & \Sigma_{\mathrm{B}}\end{array}\right),
\end{equation}
where $\Sigma_{\mathrm{A},\:\mathrm{ch},\:\mathrm{rec}}\in \mathbb{R}^{8\times8}$ is a matrix describing the Alice's mode and the state of the channel and the detection module, $\Sigma_{\mathrm{B}}\in \mathbb{R}^{2\times2}$ is a matrix describing the Bob's mode, and $\Sigma_{C}\in \mathbb{R}^{8\times2}$ is a matrix describing quadrature correlations between $\Sigma_{\mathrm{A},\:\mathrm{ch},\:\mathrm{rec}}$ and $\Sigma_{\mathrm{B}}$.

The $\Sigma_{\mathrm{tot \mid B}}$ matrix after a projective measurement of the Bob's mode depends on whether it performs homodyne or heterodyne detection.

In the case of heterodyne detection, the remaining modes are projected into the state described by the $8 \times 8$ matrix
\begin{align}
\Sigma_{\mathrm{tot} \mid \mathrm{B}}=\Sigma_{\mathrm{A},\:\mathrm{ch},\:\mathrm{rec}}-\frac{1}{V_{\mathrm{B}}+1} \Sigma_{\mathrm{C}} \Sigma_{\mathrm{C}}^{T}.
\end{align}

Again, there is no need to evaluate the entire covariance matrix. Instead, one can evaluate a block describing the eavesdropper information, that is two modes representing the entangled state that was used to model the noise and channel loss. The block itself is expressed as follows
\begin{align}
\label{eq:cov-matrix-cond}
\Sigma_{\mathrm{E \mid B}}&=\frac{1}{V_{\mathrm{B}}+1}\left(\begin{array}{ll}e_{1} \mathbf{1}_{2} & e_{2} \sigma_{z} \\ e_{2} \sigma_{z} & e_{3} \mathbf{1}_{2}\end{array}\right),\\
\notag
e_{1}&= V\left(\left(1-T_{\mathrm{rec}}\right) W_{\mathrm{rec}}+T_{\mathrm{rec}} W_{\mathrm{ch}}+1\right) +\\\label{eq:cov-matrix-1} &+T_{\mathrm{ch}}\left(W_{\mathrm{ch}}-V\right)\left(1+\left(1-T_{\mathrm{rec}}\right) W_{\mathrm{rec}}\right), \\ \label{eq:cov-matrix-2}
e_{2}&= \sqrt{T_{\mathrm{ch}}\left(W_{\mathrm{ch}}^{2}-1\right)}\left(T_{\mathrm{rec}} V+\left(1-T_{\mathrm{rec}}\right) W_{\mathrm{rec}}+1\right), \\ \label{eq:cov-matrix-3}
e_{3}&=\left(1-T_{\mathrm{rec}}\right) W_{\mathrm{ch}} W_{\mathrm{rec}}+T_{\mathrm{rec}} T_{\mathrm{ch}}\left(V W_{\mathrm{ch}}-1\right)+T_{\mathrm{rec}}+W_{\mathrm{ch}}.
\end{align}

Considering that the matrix $\Sigma_{\mathrm{E \mid B}}$ can also be represented in the form of Eq.~\eqref{eq:matrix-form}, then the symplectic eigenvalues can also be represented similarly to the formula~\eqref{eq:eigen-formula} as
\begin{align}
v_{3,\:4}&=\frac{z \pm\left(e_{3}-e_{1}\right)}{2\left(V_{B}+1\right)},\\ z&=\sqrt{\left(e_{1}+e_{3}\right)^{2}-4 e_{2}^{2}}.
\end{align}

Thus, the necessary values for estimating the Holevo bound in the presence of collective attacks in the trusted receiver noise scenario have been obtained. Further, the model will need to be supplemented taking into account the lack of access of Eve to the noise of Alice in order to consider the full scenario of trusted noise from trusted nodes.

The noise on Alice side can be composed of noise from laser power fluctuation and imperfect modulation~\cite{Jouguet2012,Laudenbach2017}. Models of such trusted Alice noise for collective attacks are presented in~\cite{Filip2008,Usenko2010,Jacobsen2015,Laudenbach2019}. This noise is modeled by analogy with the noise of the channel and the detection module using an additional thermal state $\mathrm{Th}_{\mathrm{pr}}$ with variance
\begin{equation}
W_{\mathrm{pr}}=\frac{\xi_{\mathrm{pr}}}{1-T_{\mathrm{pr}}}+1,
\end{equation}
where $T_{\mathrm{pr}}$ is a transmittance coefficient of Alice module.

Then the Alice noise measured by Bob will be $\xi_{\mathrm{pr,\:B}}=T_{\mathrm{ch}}T_{\mathrm{rec}}\xi_{\mathrm{pr}}$. The thermal state $\mathrm{Th}_{\mathrm{pr}}$ interacts with the Bob's mode through a beam splitter with a transmittance $T_{\mathrm{pr}}\rightarrow1$, since the initial signal power is assumed already at the output of Alice module. While the limit $T_{\mathrm{pr}}\rightarrow1$ would result in $W_{\mathrm{pr}}\rightarrow \infty$, the noise $\left(1-T_{\mathrm{pr}}\right) W_ {\mathrm{pr}}=\xi_{\mathrm{pr}}+1-T_{\mathrm{pr}} \rightarrow \xi_{\mathrm{pr}}$ of the mode reflected into the channel will be finite and good certain.

The problem can again be reduced to considering only two modes available to Eve, reducing the eigenvalue problem to a second degree polynomial. This allows to describe the trusted noise of Alice and Bob using simple analytical expressions.

The overall initial state now includes the thermal state responsible for the Alice's noise
\begin{equation}
\Sigma_{\mathrm{tot}, 0}=\mathrm{EPR}_{\mathrm{AB}} \oplus \mathrm{Th}_{\mathrm{pr}} \oplus \mathrm{EPR}_{\mathrm{ch}} \oplus \mathrm{Th}_{\mathrm{rec}}.
\end{equation}

Then one should redesignate the sequence of all beam splitters  $
\mathrm{BS}_{\mathrm{tot}}=\mathrm{BS}_{\mathrm{rec}} \mathrm{BS}_{\mathrm{ch}} \mathrm{BS}_{\mathrm{pr}}$. Then the sympectic transformation, analogous to Eq.~\eqref{eq:symp-bs}, will change the block of the covariance matrix related to the eavesdropper as
\begin{equation}
\label{eq:cov-mat-e-2}
\Sigma_{\mathrm{E}}=\left(\begin{array}{cc}\left(\left(1-T_{\mathrm{ch}}\right)\left(V+\xi_{\mathrm{pr}}\right)+T_{\mathrm{ch}} W_{\mathrm{ch}}\right) \mathbf{1}_{2} & \sqrt{T_{\mathrm{ch}}} \sqrt{W_{\mathrm{ch}}^{2}-1} \sigma_{z} \\ \sqrt{T_{\mathrm{ch}}} \sqrt{W_{\mathrm{ch}}^{2}-1} \sigma_{z} & W_{\mathrm{ch}} \mathbf{1}_{2}\end{array}\right).
\end{equation}

Obviously, the matrices from the expressions~\eqref{eq:cov-mat-e-1} and~\eqref{eq:cov-mat-e-2} coincide up to the replacement $V\rightarrow V+\xi_{\mathrm {pr}}$. The symplectic eigenvalues $v_1$ and $v_2$ required for calculating $S_{\mathrm{E}}$ are again obtained by the formula~\eqref{eq:eigen-formula}. To calculate $v_3$ and $v_4$ for $S_{\mathrm{E\mid B}}$, $\Sigma_{\mathrm{tot} \mid \mathrm{B}}$ must be rearranged according to the modified shared state structure.

Thus, after substitution $T_{\mathrm{pr}} = 1$ and with the accuracy of replacement $V\rightarrow V+\xi_{\mathrm{pr}}$, the Eve's covariance matrix after heterodyne detection by Bob has the form in accordance with expression~\eqref{eq:cov-matrix-cond}. The symplectic eigenvalues $v_3$ and $v_4$ are again obtained by the formulas~\eqref{eq:cov-matrix-1}-\eqref{eq:cov-matrix-3}.

The advantage of this model is that as equipment losses increase, the Holevo bound decreases faster than mutual information~\cite{Laudenbach2019}.

\section{Gaussian quantum channel modeling}
\label{sec:gaussian}
The Gaussian quantum channel, as already mentioned in this paper, is characterized by two parameters: the transmittance and excess noise. Often CV-QKD papers do not provide clarifications on the components of these characteristics~\cite{Huang2016,Zhang2020,Hosseinidehaj2020}: an analytical assessment has been carried out only for some components.

Modeling the parameters of the Gaussian channel is necessary to obtain the value of the signal-to-noise ratio, which will be maintained at a given distance.

As has already been demonstrated in the previous section, the transmittance is composite, which is why it cannot be estimated in the aggregate, which, for example, was done in~\cite{Jouguet2013a,Zhang2020}. This assumption significantly improves the performance and the amount of allowable losses in the stability analysis but remains incorrect.

So, given that the quantum channel is an optical fiber channel, the transmittance can be estimated in accordance with the well-known expression~\cite{Agrawal2012}:
\begin{align}
    T_{\mathrm{ch}}=10^{- \zeta L/10},
\end{align}
where $\zeta$ is a fiber attenuation in dB/km.

The coefficient $T_{\mathrm{det}}$ can be obtained from equipment loss and detector efficiency as:
\begin{align}
\label{eq:transm-loss}
    T_{\mathrm{det}}=\eta_{\mathrm{det}}10^{-\mathrm{losses}/10},
\end{align}
where $\mathrm{losses}$ is a cumulative losses on the Alice's equipment.

Here it should be clarified that $T_{\mathrm{det}}$ is a transmittance in the signal arm, $T_{\mathrm{det}}^{\prime}$ is a transmittance in LO arm. The latter will be necessary for estimating the excess noise. Both coefficients are calculated using the formula~\eqref{eq:transm-loss} taking into account the fact that losses in both arms are different. For example, the power of LO is expressed as
\begin{align}
    \label{eq:lo-power}
    P_{\mathrm{LO}}=T_{\mathrm{det}}^{\prime}P_{\mathrm{LO,\:A}},
\end{align}
where $P_{\mathrm{LO,\:A}}$ is a power of LO at the output of Alice.

The cumulative losses on the equipment in Bob module are presented in the Table~\ref{tab:losses}.

\begin{table}[ht]

\caption{Cumulative losses in the signal arm and in the arm of LO obtained from~\cite{db-11,db-22,db-33,db-44,db-55,db-66,db-77,db-88}}
\small
\label{tab:losses}
\begin{tabular}{|l|l|c|}
\hline
\multicolumn{1}{|c|}{Arm} & \multicolumn{1}{c|}{Name of the optical component} & Insertion loss, dB \\ \hline
\multirow{7}{*}{Signal}     & FC/APC Connectors                                                         & 3                          \\ \cline{2-3} 
                            & \mbox{CWDM}-filter                                              & 0.6                          \\ \cline{2-3} 
                            & Polarization controller                                   & 0.05                          \\ \cline{2-3} 
                            & Polarizing beam splitter                          & 0.6                          \\ \cline{2-3} 
                            & 90-degree hybrid                                      & 3                            \\ \cline{2-3} 
                            & All components                                                    & 7,25                        \\ \hline
\multirow{8}{*}{LO}         & FC/APC Connectors                                                          & 3,6                          \\ \cline{2-3} 
                            & \mbox{CWDM}-filter                                               & 0.6                          \\ \cline{2-3} 
                            & Polarization controller                                   & 0.05                          \\ \cline{2-3} 
                            & Polarizing beam splitter                            & 0.6                          \\ \cline{2-3} 
                            & Beam splitter 10/90 6                                & 0.85                         \\ \cline{2-3} 
                            & 90-degree hybrid                                      & 3                            \\ \cline{2-3} 
                            & All components                                                   & 8.7                         \\ \hline
\end{tabular}

\end{table}

In many theoretical works, the excess noise values are approximated and fixed~\cite{Hosseinidehaj2019,Hosseinidehaj2020,Pirandola2021,Pirandola2021a,Laudenbach2019}. This is motivated by the fact that in real CV-QKD systems, the excess noise, as well as the transmittance, is estimated from experimental data on the variances of quadrature operators. As far as strictly theoretical works are concerned, the substitution is necessary only for illustrative purposes. However, for a theoretical performance evaluation of the considered CV-QKD system, it is necessary to take into account various noise sources, which will be further carried out in accordance with the work~\cite{Laudenbach2017}.

Like the transmittance, excess noise is compound. In this case, the excess noise components are:
\begin{itemize}
    \item Alice module noises:
    \begin{itemize}
        \item laser power fluctuations noise;
        \item DAC noise;
    \end{itemize}
    \item channel noises:
    \begin{itemize}
        \item phase noise;
    \end{itemize}
    \item Bob module noises:
    \begin{itemize}
        \item common-mode rejection ratio (CMRR) noise;
        \item internal noise of the balanced detector;
        \item ADC noise.
    \end{itemize}
\end{itemize}

Laser power fluctuations noise contains of two components, signal and LO ones
\begin{align}
    \xi_{\mathrm{RIN},\:\mathrm{sig}}&=V_{\mathrm{A}} \sqrt{\mathrm{RIN}_{\mathrm{sig}} B_{\mathrm{sig}}},\\
    \xi_{\mathrm{RIN},\:\mathrm{LO}}&=\frac{1}{4} \mathrm{RIN}_{\mathrm{LO}} B_{\mathrm{LO}} V,\\
    \xi_{\mathrm{RIN}}&=\xi_{\mathrm{RIN},\: \mathrm{LO}}+\xi_{\mathrm{RIN},\:\mathrm{sig}}
\end{align}
where $\mathrm{RIN}_{\mathrm{sig}}$ is a relative intensity noise (RIN) of signal, $\mathrm{RIN}_{\mathrm{LO}}$ is a relative intensity noise of LO, $B_{\mathrm{sig}}$ is a signal spectrum width, and $B_{\mathrm{LO}}$ is a LO spectrum width.

It is important to note that due to the fact that both the signal and LO emit from the same laser, the spectral width and relative intensity noise for them will be the same~\cite{Laudenbach2017}, i.e. $B_{\mathrm{sig}}=B_{\mathrm{LO}}\equiv B$, $\mathrm{RIN}_{\mathrm{sig}}=\mathrm{RIN}_{\mathrm{LO} }\equiv \mathrm{RIN}$.

The excess noise caused by noise from the modulating voltage side is expressed by the inequality~\cite{Laudenbach2017}
\begin{align}
\label{eq:xi-dac}
    \xi_{\mathrm{DAC}} \leq V_{\mathrm{A}}\left(\pi \alpha \frac{\sqrt{V_{\mathrm{q}}}}{V_{\pi}}+\frac{1}{2} \pi^{2} \alpha^{2} \frac{V_{\mathrm{q}}}{V_{\pi}^{2}}\right)^{2},\\
    V_{\mathrm{q}}=\mathrm{LSB}^2/12=V_{\mathrm{FS}}^2/(12\cdot 2^{2N_{\mathrm{res}}(f_{\mathrm{rep}})}).
\end{align}
where $V_{\pi}$ is a voltage required to reverse the phase by $\pi$, $\alpha$ is a DAC gain coefficient, $V_{\mathrm{q}}$ is a converter output voltage variance, $\mathrm{\mbox{LSB}}$ is aleast significant bit, $V_{\mathrm{FS}}$ is a full-scale voltage range of the converter, and $N_{\mathrm{res}}$ is a DAC resolution.

Coherent quantum signal detection requires a well calibrated phase and frequency relationship between the signal and LO. In addition, the signal initially carries a certain level of phase noise. This phase noise, as well as relative phase shifts, can be compensated with a strong reference sent by Alice~\cite{Qi2015,Soh2015,Kleis2017,Laudenbach2019a}. The reference (or pilot) signal carries well-known phase with a fixed phase relation to the original signal pulse. Bob performs heterodyne detection of the reference signal by measuring its quadratures $q$ and $p$. It can determine the deviation from a fixed and time-constant reference phase. Any of such measured phase shift is used to appropriately correct the measured phase of the quantum signal. The remaining phase noise is then expressed as~\cite{Laudenbach2017}
\begin{equation}
    \xi_{\mathrm{PR}}=\frac{1}{2} V_{\mathrm{A}} \frac{V_{\mathrm{pt}}}{N_{\mathrm{pt}}\left\langle N_{\mathrm{pt}}\right\rangle},
\end{equation}
where $V_{\mathrm{pt}}$ is a variance of the quadrature operator of the reference signal, $N_{\mathrm{pt}}$ is a number of the reference signals, $\left\langle N_{\mathrm{pt}}\right\rangle$ is a the average number of photons in the reference pulse.

It should be noted that at purposed paper, it is assumed that for each signal message there is a reference signal, i.e. the number of reference signals is half of the total number of messages.

It is convenient to express the internal noise of a balanced detector in terms of a characteristic called clearance $C$, which is defined as the ratio of the total experimental variance of a zero power signal (dispersion shot noise~$V_{0}(\hat{q })$ and electronic noise of dispersion~$V_{\operatorname{det}}(\hat{q})$) and variance~$V_{\operatorname{det}}(\hat{q})$ caused only by detector electronic noise:
\begin{align}
\label{eq:clearance}
C=\frac{V_{0}(\hat{q})+V_{\operatorname{det}}(\hat{q})}{V_{\operatorname{det}}(\hat{q})}.
\end{align}

The shot noise variance depends linearly on the power of LO, which, however, is limited by the saturation limit of the detector's PIN diodes. Experimentally, the numerator of the Eq.~\eqref{eq:clearance} can be determined by measuring the quadrature variance of LO when it is mixed with the vacuum inlet. The denominator is the remaining quadrature variance after LO is disconnected from the detector. In SNU by definition $V_{0}(\hat{q})=1$, and the equation becomes
\begin{align}
C=\frac{1+V_{\mathrm{det}}(\hat{q})}{V_{\mathrm{det}}(\hat{q})}=\frac{1+\xi_{\mathrm{det}}}{\xi_{\mathrm{det}}}.
\end{align}

Thus, the noise of a balanced detector relative to the clearance, taking into account one/two detectors in homo-/heterodyne detection respectively, is
\begin{align}
\xi_{\mathrm{det}}=\mu \frac{1}{C-1}.
\end{align}

\begin{figure}[ht!]
\centering
\includegraphics[width=0.8\textwidth]{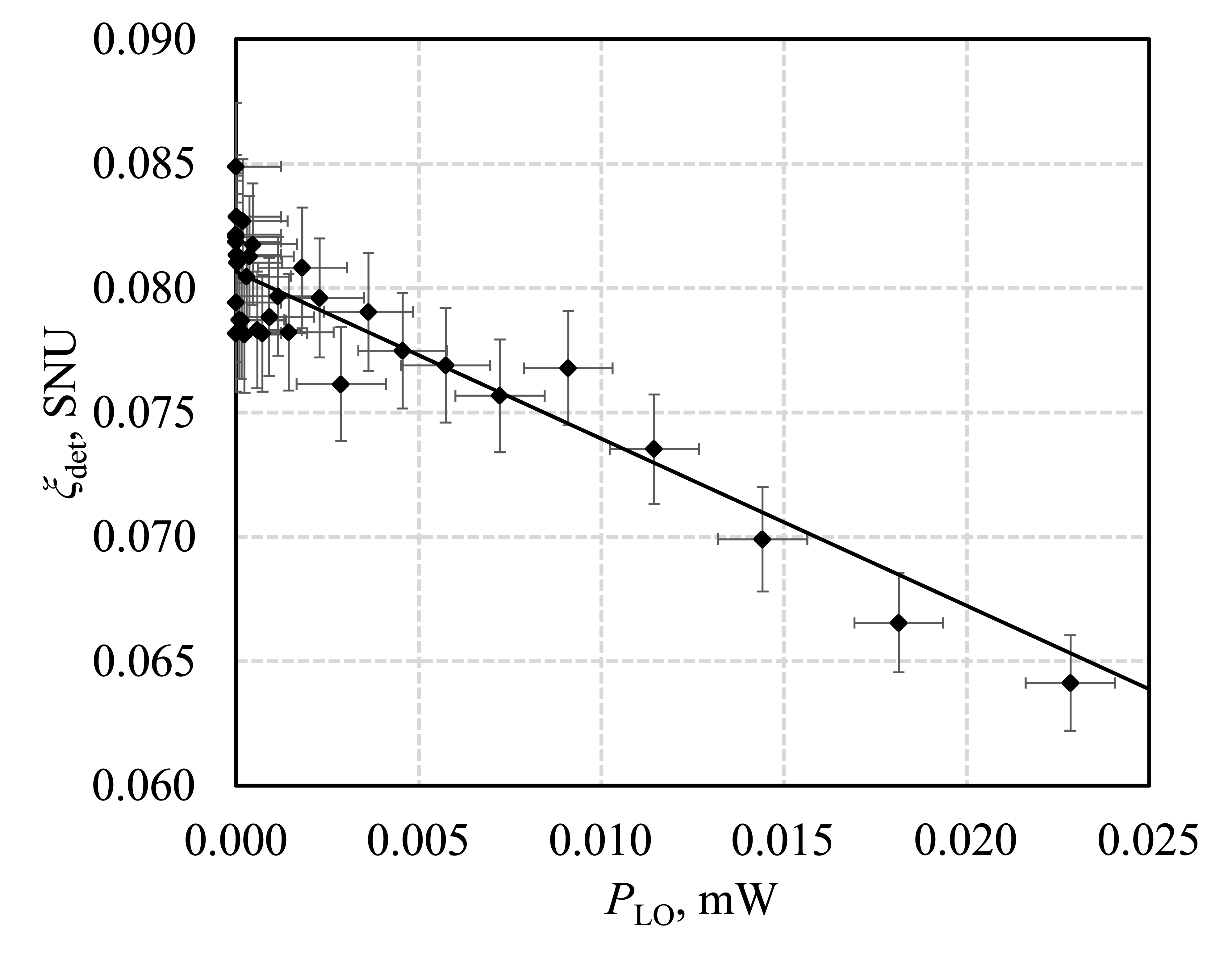}
\caption{Dependence of excess noise of the balanced detector on the input power of LO for General Photonics OEM Balanced Detector (BPD-003).}  
\label{fig:xi-det}
\end{figure}

An experimental evaluation of $\xi_{\mathrm{det}}$ was carried out for the General Photonics OEM Balanced Detector (BPD-003) with an operating frequency band of 200~MHz (see Figure~\ref{fig:xi-det}). The choice of the detector was based on the following parameters for the optimal signal-to-noise ratio:
\begin{itemize}
    \item low noise equivalent power~--- indicates the possibility of detecting signals with a power comparable to that of shot noise;
    \item high gain coefficient~--- allows detection of low power signals with high attenuation of LO;
    \item high CMRR~--- shows the gain quality;
    \item a wide operating frequency band~--- allows to increase the frequency of sending states, which, in accordance with the current technical level, should be about MHz~\cite{Jouguet2013a,Huang2016,Zhang2020}. The established relationship between the excess noise of the detector and its operating frequency band is linear~\cite{Laudenbach2017}, so a too large range can lead to a high level of internal detector noise.
\end{itemize}

The dependence of excess noise of the detector and, as a result, the signal-to-noise ratio on the operating frequency band does not have a minimum due to linearity, however, as can be seen in Figure~\ref{fig:xi-det}, the obtained noise level in the considered detector is quite low and corresponds to the level established by~\cite{Laudenbach2017,Zhang2020}. So, for example, with a total loss of 13.7 dB (5 dB in the channel and 8.7~dB in LO arm), $\xi_{\mathrm{det}}=0.093$ is observed in~\cite{Laudenbach2017} at operating frequency band of the detector is 250~MHz. In~\cite{Zhang2020} $\xi_{\mathrm{det}}\sim 10^{-1}$, based on the given data.

According to the obtained dependence of the excess noise of the balanced detector on the input power of LO, one can observe an increase in the contribution of the noise with a decrease in its values. In practice, a decrease in power is associated with an increase in losses. The resulting dependence will be used later in the overall assessment of the total excess noise. The experimental background of the obtained formula is related to the fact that the proposed models do not take into account the full composition of the balanced detector.

A realistic differential amplifier will amplify not only the difference current with a coefficient $g$, but also to a small extent with a coefficient $g_{\mathrm{CM}}$ their average value of the input photocurrents in the subtractive circuit. As a characteristic for estimating such amplification, CMRR~\cite{Szynowski1983} is used:
\begin{align}
    \mathrm{CMRR}&=\left|\frac{g}{g_{\mathrm{CM}}}\right|.
\end{align}

Noise dependent on CMRR is calculated as~\cite{Laudenbach2017}:
\begin{align}
    \xi_{\mathrm{CMRR}}=\frac{\mu}{4 \mathrm{CMRR}^{2}}\left(\frac{h f V_{\mathrm{A}}^{2}}{4 \tau P_{\mathrm{LO}}} \mathrm{RIN}_{\mathrm{sig}} B_{\mathrm{sig}}+\frac{\tau}{h f} P_{\mathrm{LO}} \mathrm{RIN}_{\mathrm{LO}} B_{\mathrm{LO}}\right),
\end{align}
where $P_{\mathrm{LO}}$ is a LO power, $\tau$ is a pulse duration, $f$ is an optical frequency and $h$ is a Planck's constant.

The incoming signal pulse will be received and amplified by Bob's balanced detector, where the output voltage will be proportional to the measured quadrature. However, if the output voltage is quantized by DAC, it will introduce additional error into the measured signal, thereby contributing to excess noise as~\cite{Laudenbach2017}:
\begin{equation}
    \label{eq:xi-adc}
    \xi_{\mathrm{ADC}}=\mu \frac{\tau V_{\mathrm{q}}}{h f \mathrm{~g}^{2} \rho^{2} P_{\mathrm{LO}}}
\end{equation}
where $g$ is a gain coefficient of the electrical circuit of balanced detector and $\rho$ is a photodiode responsivity.

It should be noted that in the equations~\eqref{eq:xi-dac} and~\eqref{eq:xi-adc} the output voltage variance of DAC and ADC are equal, because they are selected with the same scope and resolution.

The calculation of the components of the total excess noise was carried out in accordance with the parameters specified in Table~\ref{tab:parameters}. The substantiation of the variance of the quadrature operator is given below in Sec.~\ref{sec:mean-phot}. It is important that in the considered case of transmitted LO (see Figure~\ref{fig:gen-scheme}), the power of the LO, as well as the signal power, depends significantly on the losses on equipment and in the channel, which must be taken into account in assessing the excess noise related to the receiver.
\begin{longtable}[c]{|c|c|c|c|}
\caption{Parameters used in modeling excess noise obtained from Refs.~\cite{Laudenbach2017,Tang2020}.}
\label{tab:parameters}
\\
\hline
\begin{tabular}[c]{@{}c@{}}Parameter\end{tabular} & Description                                                                                                & Value              & \begin{tabular}[c]{@{}c@{}}Units\end{tabular} \\ \hline
\endfirsthead
\multicolumn{4}{@{}l}%
{{Continuation of Table \thetable}} \\
\hline
\begin{tabular}[c]{@{}c@{}}Parameter\end{tabular} & Description                                                                                                & Value              & \begin{tabular}[c]{@{}c@{}}Units\end{tabular} \\ \hline
\endhead
$V_{\mathrm{A}}$                                                 & \begin{tabular}[c]{@{}c@{}}Alice's modulation\\variance\end{tabular}                               & 6.77                  & SNU                                                        \\ \hline
$\mathrm{RIN}$                                                   & \begin{tabular}[c]{@{}c@{}}RIN\end{tabular}                              & $10^{-14.5}$          & Hz$^{-1}$                                                   \\ \hline
$B$                                                              & laser spectrum width                                                                                   & $10^4$                & Hz                                                          \\ \hline
$V_{\pi}$                                                        & \begin{tabular}[c]{@{}c@{}}voltage required to \\ reverse the phase by $\pi$\end{tabular}            & 5                   & V                                                           \\ \hline
$\alpha$                                                         & DAC gain                                                                             & 8                     & a.u.                                                          \\ \hline
$V_{\mathrm{q}}$                                                 & \begin{tabular}[c]{@{}c@{}}variance of output voltage of converter\end{tabular}                & 1.94$\cdot 10^{-11}$  & V                                                           \\ \hline
$V_{\mathrm{FS}}$                                                &  full-scale voltage range of DAC                                                                                           & 1                     & V                                                           \\ \hline
$N_{\mathrm{res}}$                                               & DAC resolution                                                                                        & 16                    & bit                                                         \\ \hline
$V_{\mathrm{pt}}$                                                & \begin{tabular}[c]{@{}c@{}}variance of pilot signal\end{tabular}                          & 1.2                   & SNU                                                        \\ \hline
$n_{\mathrm{pt}}$                                                & number of pilot signals                                                                             & $3\cdot10^{8}$       & --                                                          \\ \hline
$\left\langle N_{\mathrm{pt}}\right\rangle$                      & \begin{tabular}[c]{@{}c@{}}mean photon number in pilot puls\end{tabular}                      & 600                   & --                                                          \\ \hline
$\mu$                                                            & \begin{tabular}[c]{@{}c@{}}homo-/heterodyning parameter\end{tabular}                              & 2                     & --                                                          \\ \hline
$P_{\mathrm{LO},\:A}$                                            & \begin{tabular}[c]{@{}c@{}}Alice's output LO power\end{tabular}                  & 2$\cdot10^{-3}$       & W                                                          \\ \hline
$T_{\mathrm{det}}$                                               & \begin{tabular}[c]{@{}c@{}}transmittance of\\ signal arm\end{tabular}                    & $10^{-0,745}$                 & a.u.                                                          \\ \hline
$T_{\mathrm{det}}^{\prime}$                                      & \begin{tabular}[c]{@{}c@{}}transmittance of \\ LO arm\end{tabular}                           & $10^{-0,89}$                  & a.u.                                                          \\ \hline
$\tau$                                                           & pulse duration                                                                                & 3$\cdot10^{-9}$       & s                                                           \\ \hline
$f$                                                              & optical frequency                                                                                     & 1.934 $\cdot 10^{14}$ & Hz                                                          \\ \hline
$h$                                                              & Planck's constant                                                                                      & 6.63$\cdot 10^{-34}$  & J$\cdot$s                                                  \\ \hline
$g$                                                              & \begin{tabular}[c]{@{}c@{}}balanced detector's gain\end{tabular} & 10$^5$                & V/A                                                         \\ \hline
$\rho$                                                           & photodiode responsivity                                                                            & 0.85                  & A/W                                                        \\ \hline
$\mathrm{CMRR}$                                                  & CMRR                                                                                                  & 30                    & dB                                                          \\ \hline
\end{longtable}

\section{Evaluation and monitoring of experimental parameters}
\label{sec:parameters}
This subsection describes the procedure for assessing, optimizing and monitoring the key parameters of the CV-QKD. Since the selection of parameters directly affects the secure key generation rate, it is necessary to introduce a boundary with respect to which the calculation will be carried out.

In the general case, the secure key generation rate $K$ is determined by
\begin{equation}
K=f_{\mathrm{sym}} \cdot r,
\end{equation}
where $f_{\mathrm{sym}}$ is a repetition rate and $r$ is a secure key fraction.

The asymptotic secure key generation rate in terms of a message with ideal post-processing for the CV-QKD system in the case of collective attacks is given by the Devetak-Winter bound~\cite{Devetak2005} as
\begin{equation}
\label{eq:devetak}
r_{\mathrm{coll}}^{\mathrm{asympt}} \geqslant I_{\mathrm{AB}}-\chi.
\end{equation}

Given the non-ideal reverse reconciliation, the bound can be refined as
\begin{equation}
\label{eq:r-asympt}
r_{\mathrm {coll }}^{\mathrm {asympt }} \geqslant(1-\mathrm{FER})\left(\beta I_{\mathrm{AB}}-\chi_{\mathrm{E B}}\right),
\end{equation}
where $\mathrm{FER}\in [0,\:1]$ is a frame error rate (FER).

\begin{figure}[ht]
\centering
\includegraphics[width=0.7\textwidth]{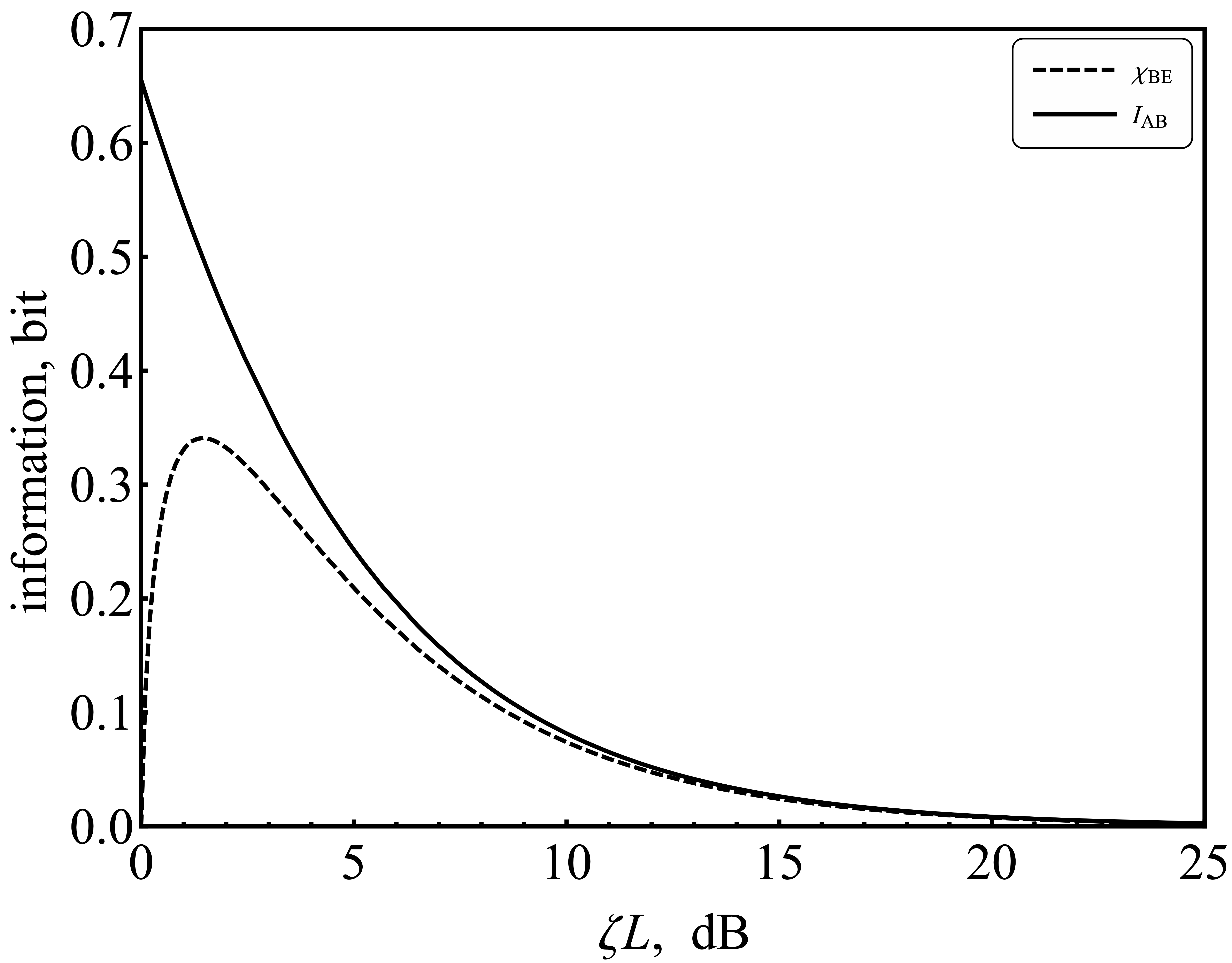}
\caption{Mutual information $I_{\mathrm{AB}}$ and the Holevo bound $\chi_{\mathrm{BE}}$ versus losses in the quantum channel.} 
\label{fig:iab-chi}
\end{figure}

The quantities $I_{\mathrm{AB}}$ and $\chi_{\mathrm{E B}}$ are respectively obtained by the formulas~\eqref{eq:mutual-inf} and~\eqref{eq:chi-final}. The dependence of these quantities on losses in the quantum channel is shown in Figure~\ref{fig:iab-chi}.

It can be seen that the Holevo bound does not exceed the mutual information at large distances, thereby keeping the secure key generation rate positive (see Figure~\ref{fig:iab-chi}). Such an assessment poorly reflects reality, because with an infinite number of messages, and, hence, with an infinite sample for estimating the parameters and an infinite number of reference signals, the excess noise in the channel is very small. At the same time, the excess noise and losses on Bob side are still large. Limitations on the quantum channel length, as, for example, in~\cite{Laudenbach2019}, are due to the fact that the excess noise model uses not analytical expressions, but approximate fixed values. It can also be noted that when using error correction codes over an alphabet of size $q$, each symbol of the code corresponds to $\log_{2} q$ bits, and for a given code rate $R$, which is selected based on the channel parameters, one can write~ \cite{Laudenbach2017}:
\begin{equation}
  R \log_{2} q=\beta I_{\mathrm{AB}}.
\end{equation}
\begin{figure}[ht]
\centering
\includegraphics[width=0.7\textwidth]{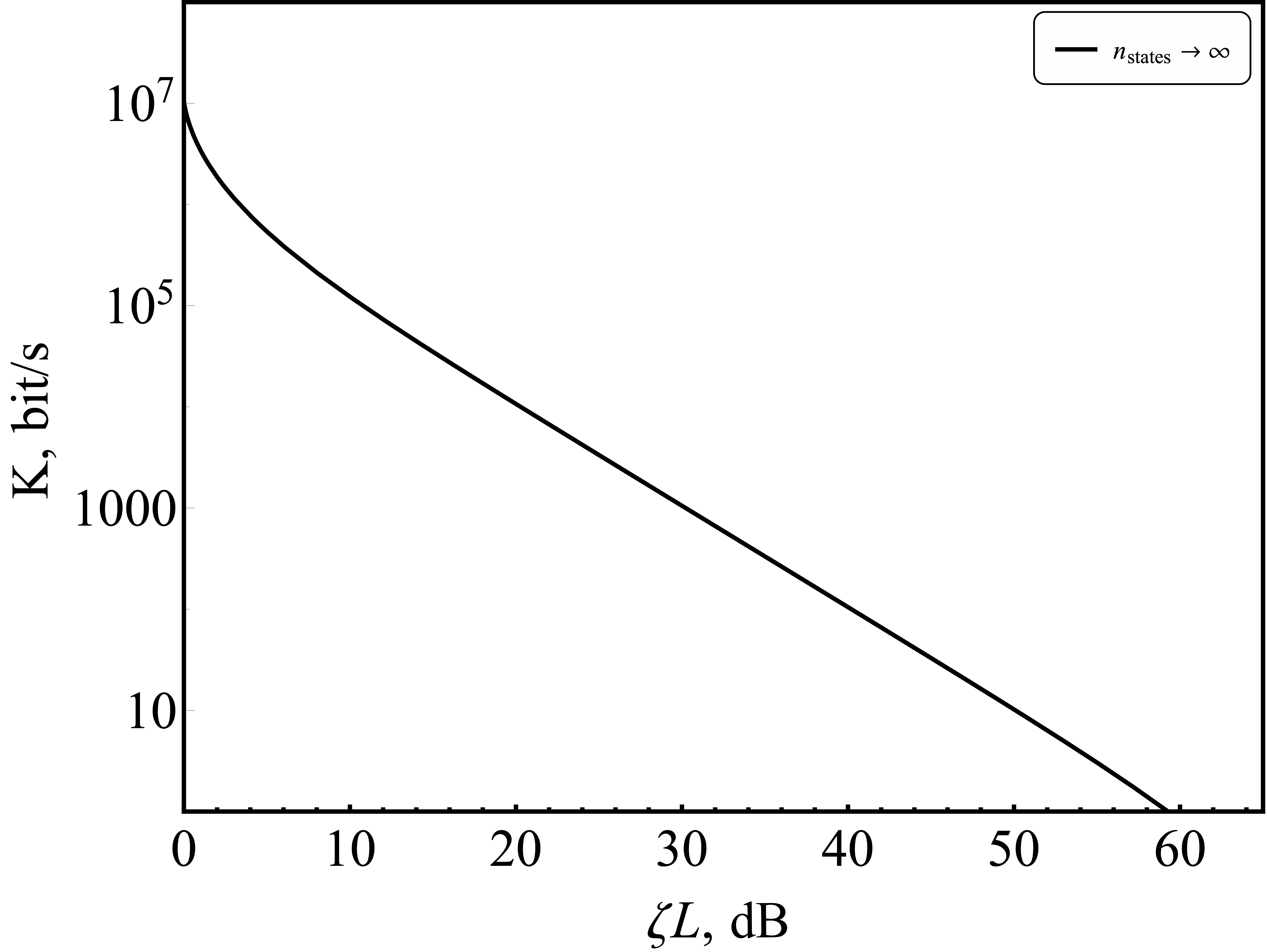}
\caption{Dependence of secure key generation rate in the asymptotic limit on losses in the quantum channel considering the presence of collective attacks.}  
\label{fig:gmcs-col-key-asymp}
\end{figure}

\subsection{Mean photon number optimization}
\label{sec:mean-phot}
Obviously, an increase in the variance of the Alice's quadrature operator $V_{\mathrm{A}}$, which specifies the average number of photons over an ensemble of states (see Eq.~\eqref{eq:mean-phot-fin}), will lead to a proportional increase in the signal-to-noise ratio (see Eq.~\eqref{eq:snr}). That is, an increase in the average number of photons in the signal will give better discrimination of Gaussian states up to the limits provided by the detector. However, it must be taken into account that in this case Eve will also receive more information. In this regard, it is necessary to estimate the value of $V_{\mathrm{A}}$ by maximizing the value of the secure key fraction $r$ at the target distance.

Thus, with a target loss in the quantum channel of 5~dB and the efficiency of the reconciliation procedure $\beta=0.95$, the optimal value $V_{\mathrm{A}}$ is 6.77 SNU, which is obtained by maximizing the adjusted value of $r$ from the expression~\eqref{eq:key-compos}. It is important to note that this parameter is significantly affected by the total number of states $n_{\mathrm{states}}$, which, in turn, determines the limit for the number of reference signals.

\subsection{Parameter estimation}
The Bob measurement data does not initially correspond to Gaussian quantum information terminology. For this reason, it is necessary to introduce a conversion coefficient $\phi$, expressed in V$^2$/SNU, which converts the original data to SNU. To keep track of possible changes to the calibration parameters, this procedure should be repeated during the key exchange step. Instead of the variance of the quadrature operator, Bob measures the voltage variance (see Eq.~\eqref{eq:var-bob})
\begin{equation}
V(U)=\phi V\left(\hat{q}_{\mathrm{B}}\right).
\end{equation}

Alice and Bob randomly jointly select $n_{\mathrm{pe}}$ from $n_{\mathrm{states}}$ distributed signals and publicly disclose the corresponding $\mu n_{\mathrm{pe}}$ value pairs. Under the collective Gaussian attack assumption, these pairs are independent and equally distributed Gaussian variables. In accordance with the maximum likelihood method, the following estimate can be obtained from the sample
\begin{equation}
V(U)=\langle U^{2}\rangle-\langle U\rangle^{2}=\frac{1}{\mu n_{\mathrm{pe}}} \sum_{i=1}^{\mu n_{\mathrm{pe}}} U_{i}^{2}-\left(\frac{1}{\mu n_{\mathrm{pe}}} \sum_{i=1}^{\mu n_{\mathrm{pe}}} U_{i}\right)^{2},
\end{equation}
where $U_{i}$ is a measured voltage value.

Approximately the parameter $\phi$ can be estimated as
\begin{equation}
\phi \approx P_{\mathrm{LO}} \rho^{2} g^{2} B_{\mathrm{BD}} h f,
\end{equation}
where $B_{\mathrm{BD}}$ is an operating frequency of the balanced detector.

However, the coefficient $\phi$ must be determined experimentally more precisely. To do this, Bob disables the signal input ($TV_{\mathrm{A}} = \xi_{\mathrm{ch}} = 0$) and instead measures the quadratures of the vacuum state. Then the variance of Bob quadrature operator is
\begin{align}
V\left(\hat{q}_{B}\right)&=1+\frac{\xi_{\mathrm{rec}}}{\mu},\\
\label{eq:vu}
V(U) &=\phi+\phi \frac{\xi_{\mathrm{rec}}}{\mu} \equiv \phi+N_{\mathrm{rec}}.
\end{align}

It should be noted that $\phi$ is linearly directly proportional to LO power, while the detector noise is inversely proportional, as shown in the expression~\eqref{eq:xi-adc} and as can be seen from Figure~\eqref{fig:xi-det} (the analytical formula for the detector noise is presented in~\cite{Laudenbach2017}). Therefore, the product of two is constant with respect to $P_{\mathrm{LO}}$
\begin{align}
\phi &\propto P_{\mathrm{LO}}, \\ \xi_{\mathrm{rec}} &\propto \frac{1}{P_{\mathrm{LO}}}, \\ \frac{\partial N_{\mathrm{rec}}}{\partial P_{\mathrm{LO}}}&=\frac{\partial\left(\phi \xi_{\mathrm{rec}}\right)}{\partial P_{\mathrm{LO}}}=0.\end{align}

Thus, in the case of $P_{\mathrm{LO}} = 0$, the coefficient $\phi$ will become zero, but $N_{\mathrm{rec}}$ will remain unchanged, since it does not depend on LO power. Therefore, when not only the signal but also LO input are disabled, the expression~\eqref{eq:vu} becomes
\begin{equation}
V(U)=N_{\mathrm{rec}}.
\end{equation}

Now, for a given voltage dispersion $V (U)$, obtained with a given non-zero LO power, we can write the final formula for $\phi$:
\begin{equation}
\phi=V(U)-N_{\mathrm{rec}}.
\end{equation}

The quantity $\phi$ is a quadratic measure of the voltage of exactly one SNU, still assuming that only the vacuum input is measured, i.e. $T V_{\mathrm{A}} = 0$. For subsequent parameter estimation, Bob divides his measured voltages representing $q$ and $p$ by $\sqrt{\phi}$ and any calculated voltage variance by $\phi$, so that all his data will be represented in SNU system.

\subsection{Confidence intervals}
Depending on the chosen security model (trusted or untrusted noise), when estimating the parameters, it is necessary to set certain confidence intervals. Since the noise of the equipment is assumed to be trusted in the model under consideration, the total transmittance for Bob (in the expression~\eqref{eq:lo-power}) must be estimated in terms of the best case, while the transmittance and excess noise of the channel~--- in terms of the worst.

As already mentioned, Alice and Bob have a sample of $\mu n_{\mathrm{pe}}$ independent and equally distributed pairs $\{x_i,\:y_i\}_{i=1}^{\mu n_{\ mathrm{pe}}}$, where $x_i$ and $y_i$ are Gaussian variables, which are related by the ratio of the channel with additive white Gaussian noise:
\begin{align}
    y=\sqrt{T}x+\mathcal{N}(0,\:\xi).
\end{align}

For pairs of values, estimates of the transmittance and excess noise are determined:
\begin{align}
    \hat{T}&=\frac{\sum_{i=1}^{\mu n_{\mathrm{pe}}}x_{i}y_{i}}{\sum_{i=1}^{\mu n_{\mathrm{pe}}}x_{i}^2},\\
    \hat{\xi}&=\frac{1}{\mu n_{\mathrm{pe}}}\sum_{i=1}^{\mu n_{\mathrm{pe}}}(y_{i}-\hat{T}x_{i}).
\end{align}

It should be taken into account that corrections must be introduced, depending on the belonging of the noise for a correct assessment of the mutual information and the Holevo bound. At the same time, the channel model with additive noise is preserved. The corrections themselves are expressed as~\cite{Pirandola2021a}:
\begin{align}
    \operatorname{Corr}_{\xi,\:j}&=w\sqrt{\operatorname{Var}(\hat{T}_{j}^{1/2})}=w \frac{\xi_{j}+\mu}{\sqrt{2\mu n_{\mathrm{pe}}}},\\
    \label{eq:confid-T}
    \operatorname{Corr}_{T,\:j}&=w\sqrt{\operatorname{Var}(\hat{\xi}_{j})}=2w\sqrt{\frac{2 T_{j}^2+T_{j}(\xi_{j}+\mu)/V_{\mathrm{A}}}{\mu n_{\mathrm{pe}}}},
\end{align}
where $w$ is a confidence factor, $m$ is a number of signals for parameter estimation and $j$ is a parameter that defines belonging to Alice/channel/Bob.

Such approximations are correct up to $O(n_{\mathrm{pe}}^{-1})$. The expression $\operatorname{Var}(\hat{T}^{1/2})$ can be further approximated for large $n_{\mathrm{pe}}$, so a more optimistic estimate can be written
\begin{align}
    \operatorname{Corr}_{T,\:j}&=2w\xi_j/V_{\mathrm{A}}\sqrt{T_{j}/(\mu n_{\mathrm{pe}})}.
\end{align}
However, the estimate from the expression~\eqref{eq:confid-T} will be used for performance analysis.

Thus, it is necessary to carry out the replacement as follows
\begin{align}
    T_{\mathrm{det}}^{\prime} &\longrightarrow T_{\mathrm{det}}^{\prime}+\operatorname{Corr}_{T,\:\mathrm{full}},\\
    T_{\mathrm{ch}} &\longrightarrow T_{\mathrm{ch}}-\operatorname{Corr}_{T,\:\mathrm{ch}},\\
    \xi_{\mathrm{ch}} &\longrightarrow \xi_{\mathrm{ch}}+\operatorname{Corr}_{\xi,\:\mathrm{ch}}.
\end{align}
Each of these estimates limits the corresponding actual value to within the error probability $\varepsilon_{\mathrm{pe}}$, if denote
\begin{equation}
\label{eq:condidence}
w=\frac{\sqrt{2}}{\operatorname{erf}\left(1-2 \varepsilon_{\mathrm{pe}}\right)}\approx \sqrt{2 \ln \left(1 / \varepsilon_{\mathrm{pe}}\right)}.
\end{equation}

Approximation in Eq.~\eqref{eq:condidence} is allowed for small $\varepsilon_{\mathrm{pe}}\leq 10^{-17}$.

\section{Estimating the finite-length secure key generation rate}
\label{sec:security}
After parameter estimation, each initial sequence of $n_{\mathrm{states}}$ dimensions goes into $n$ symbols to be processed into the final key using error correction and privacy amplification procedures. For each information block, errors are successfully corrected with a probability of $1-\mathrm{FER}$. The value of this probability depends on the signal-to-noise ratio, the target reconciliation efficiency $\beta$, and the $\varepsilon$-criteria correctness $\varepsilon_{\mathrm{cor}}$. The latter limits the probability that local bitstrings of Alice and Bob are different after error correction and successful execution of the validation procedure.

On average, $n(1-\mathrm{FER})$ signals from the information block remain for the privacy amplification procedure. This final step is implemented with the $\varepsilon$-security parameter $\varepsilon_{\mathrm{sec}}$, which limits the trace distance between the final key and the ideal key, which has no correlation with the eavesdropper. In the QKD paradigm, it is necessary to take into account the pessimistic assessment of information distributed among users. In this case not Shannon entropies, but smoothed Rényi min-entropies are used, which are reduced to the former through the asymptotic equipartition property~\cite{Tomamichel2009}. The smoothness determines the allowable error fluctuations. In turn, $\varepsilon$-security is technically decomposed as:
\begin{align}
    \varepsilon_{\mathrm{sec}}=\varepsilon_{\mathrm{s}}+\varepsilon_{\mathrm{h}},
\end{align}
where $\varepsilon_{\mathrm{s}}$ is a min-entropy smoothing parameter and $\varepsilon_{\mathrm{h}}$ is a parameter that determines the match of hash codes after privacy amplification procedure.

All declared $\varepsilon$-security parameters are set small (for example, $2^{-33} \approx 10^{-10}$) and form a general security criteria
\begin{align}
    \varepsilon=2(1-\mathrm{FER})\varepsilon_{\mathrm{pe}}+\varepsilon_{\mathrm{cor}}+\varepsilon_{\mathrm{sec}}.
\end{align}

\subsection{Satisfying the composability criteria of CV-QKD protocol in the presence of collective attacks}
\label{sec:col-att}
Taking into account the finiteness of the keys and the requirement that the protocol under consideration be composable in the presence of collective attacks, the boundary from the expression~\eqref{eq:r-asympt} is refined, and the secure key generation rate in terms of the message is expressed as~\cite{Pirandola2021,Pirandola2021a}
\begin{align}
\label{eq:key-compos}
r_{\mathrm {coll }}^{\mathrm {finite }} &\geqslant \frac{n (1-\mathrm{FER})}{n_{\mathrm{states}}}\left(\beta I_{\mathrm{AB}}(w)-\chi_{\mathrm{E B}}(w)-\frac{\Delta_{\mathrm{AEP}}}{\sqrt{n}}+\frac{\Theta}{n}\right),\\
\Delta_{\mathrm{AEP}}&=4 \log _{2}(2 \sqrt{d}+1) \sqrt{\log _{2}\left(\frac{18}{(1-\mathrm{FER})^{2} \varepsilon_{\mathrm{s}}^{4}}\right)},\\
\Theta&=\log _{2}\left[(1-\mathrm{FER})\left(1-\varepsilon_{\mathrm{s}}^{2} / 3\right)\right]+2 \log _{2} \sqrt{2} \varepsilon_{\mathrm{h}},
\end{align}
where $n$ is a number of characters left to process the final key, $n_{\mathrm{states}}$ is the number of states, $\Delta_{\mathrm{AEP}}$ is a correction according to asymptotic equipartition property~\cite{Tomamichel2009}, $\Theta$ is a correction coefficient that combines hash mismatch accounting after privacy amplification procedure according to Lemma 2 of~\cite{Tomamichel2011} and a leak on the error correction procedure~\cite{Pirandola2021,Pirandola2021a}, $d$ is a size of the effective alphabet after the final digitization of the continuous variables of Alice and Bob and $\varepsilon_{j}$ is the security parameter.

To improve performance, $\Delta_{\mathrm{AEP}}$ can be refined as~\cite{Pirandola2021a,Tomamichel2016}
\begin{align}
    \Delta_{\mathrm{AEP}}&=4 \log _{2}(\sqrt{d}+2) \sqrt{\log _{2}\left(\frac{18}{(1-\mathrm{FER})^{2} \varepsilon_{\mathrm{s}}^{4}}\right)}.
\end{align}

The value of $n$, in turn, is obtained from $n_{\mathrm{states}}$ as follows
\begin{align}
    n=n_{\mathrm{states}}-(n_{\mathrm{pt}}+n_{\mathrm{pe}}),
\end{align}
where $n_{\mathrm{pt}}$ --- number of reference signals and $n_{\mathrm{pe}}$ --- number of signals given for parameter estimation.

\subsection{Security of CV-QKD protocol against coherent attacks}
So far, the security of the CV-QKD protocol with Gaussian modulation in the presence of Gaussian collective attacks has been substantiated. The level of security for a protocol with heterodyne detection can be extended to security against coherent attacks using the mathematical apparatus from~\cite{Leverrier2017}.

Let the protocol $\mathcal{P}$, which uses coherent states as information carriers, be $\varepsilon$-secure with a secure key generation rate of finite length $r_{\mathrm {coll }}^{\mathrm {finite }}$ in the presence of collective Gaussian attacks, and $\mathcal{P}$ can be symmetrized with respect to the representation of the group of unitary matrices in the Fock space. This symmetrization is equivalent to applying an identical random orthogonal matrix to the classical continuous variables~\cite{Leverrier2017}, which is certainly possible for a protocol that implies heterodyne detection. The symmetrized protocol can be denoted as $\tilde{\mathcal{P}}$.

Then it can be assumed that users jointly perform the so-called energy tests on the sample $n_{\mathrm{et}} = f_{\mathrm{et}}n$ from random inputs for some coefficient $f_{\mathrm{et}}<1$. In each test, the parties measure a local average of the number of photons, which can be extrapolated from the data, and calculate the average over the $n_{\mathrm{et}}$ tests. If these averages exceed the specified thresholds ($d_{\mathrm{A}}$ for Alice and $d_{\mathrm{B}}$ for Bob), the protocol is aborted. Setting $d_{\mathrm{A}} \geqslant V_{\mathrm{A}}/2+O(n_{\mathrm{et}})$ guarantees almost successful passing of the test with probability $p_{\mathrm{et }}\approx1$ in typical scenarios, where the signals are attenuated and the noise is not too high, at large values of $n_{\mathrm{et}}$~\cite{Pirandola2021}. Also, for a channel with losses and a sufficiently small excess noise, the average number of photons reaches Bob, which is clearly less than in the state prepared by Alice, which means that a successful value for $d_{\mathrm{B}}$ can be chosen to be $d_{\mathrm{A}}$, i.e. relies $d_{\mathrm{A}}=d_{\mathrm{B}}\equiv d_{\mathrm{et}}$.

Thus, the parties are moving to a symmetrized $\tilde{\mathcal{P}}$ protocol, which will now use $\tilde{n}=n_{\mathrm{states}}-n_{\mathrm{coh}}$ signals to generate secret quantum keys, where $n_{\mathrm{coh}}\equiv n_{\mathrm{pt}}+n_{\mathrm{pe}}+n_{\mathrm{et}}$.

Moreover, additional privacy amplification is required, reducing the output key string by ~\cite{Leverrier2017,Pirandola2021,Pirandola2021a} $\Phi_n$
\begin{align}\Phi_{n}&=2\left[\log _{2}\left(\begin{array}{c}K_{n}+4 \\ 4\end{array}\right)\right], \\ K_{n}&=\max \left\{1,2 \tilde{n} d_{\mathrm{et}} \frac{1+2 \sqrt{\vartheta}+2 \vartheta}{1-2 \sqrt{\vartheta / f_{\mathrm{et}}}}\right\},\\
\vartheta&=(2 \tilde{n})^{-1} \ln (8 / \varepsilon).
\end{align}
Assuming that the original protocol has $\varepsilon$ security criteria against collective Gaussian attacks, otherwise the security criteria for the symmetrized protocol against coherent attacks goes to~\cite{Leverrier2017}
\begin{align}
\varepsilon^{\prime}=K_{n}^{4} \varepsilon / 50.
\end{align}

It should be noted that a very strict limitation on $\varepsilon$-parameters is implied. In particular, this means that $\varepsilon_{\mathrm{pe}}$ must be sufficiently small (for example, $10^{-43}$, as suggested by~\cite{Hosseinidehaj2020,Pirandola2021}), and the corresponding coefficient confidence $w$ must be calculated using Eq.~\eqref{eq:condidence}.

Given the changed length of the input sequence and the change in the security criteria, Eq.~\eqref{eq:key-compos} is rewritten as
\begin{align}
    r_{\mathrm {coh }}^{\mathrm {finite }} &\geqslant \frac{\tilde{n}(1-\mathrm{FER})}{n_{\mathrm{states}}} \left(\beta I_{\mathrm{AB}}(w)-\chi_{\mathrm{E B}}(w)-\frac{\Delta_{\mathrm{AEP}}}{\sqrt{n}}+\frac{\Theta-\Phi_{n}}{n}\right).
\end{align}

\subsection{Analysis of the potential performance of CV-QKD system}
\begin{figure}[ht]
\centering
\includegraphics[width=0.8\textwidth]{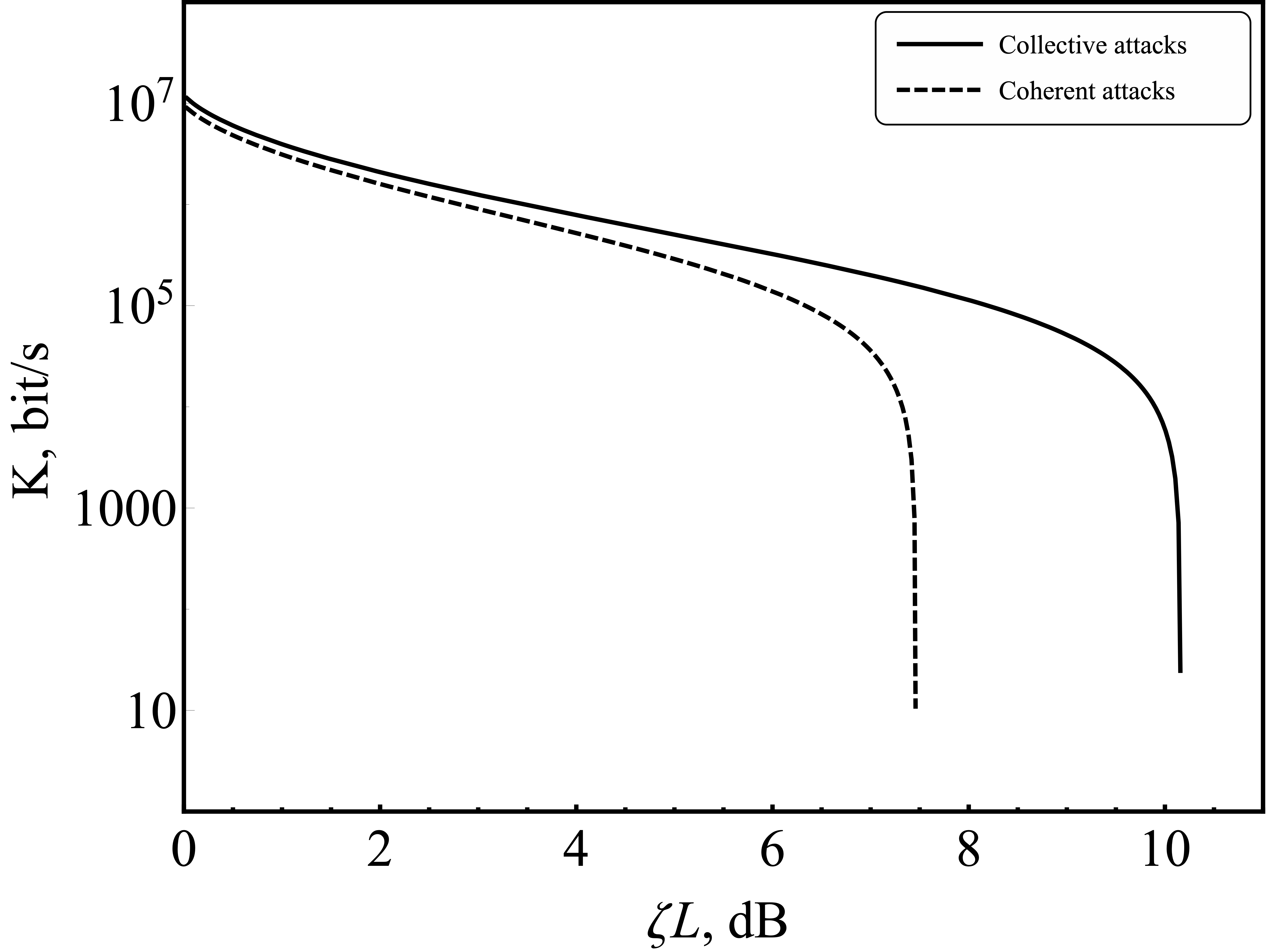}
\caption{Dependence of the secure key generation rate of finite length in the presence of collective and coherent attacks on losses in the quantum channel.}  
\label{fig:gmcs-col-coh-key}
\end{figure}

The dependence of the finite-length secure key generation rate in the presence of collective and coherent attacks on losses in the quantum channel is shown in Figure~\ref{fig:gmcs-col-coh-key}. The corresponding parameters are presented in the tables~\ref{tab:parameters} and~\ref{tab:parameters-2}. A significant contribution to the performance of any system CV-QKD is made by the number of states, while increasing this parameter imposes a limit on the computing resource. In the case under consideration, this value was estimated from the calculation of memory characteristics and information processing speed, i.e. the number of states was chosen as large as possible to fully record information about them in high-speed memory of the type DDR. It is supposed to use the Kria K26 computing module from Xilinx with a memory of 4~GB, of which 2~GB is allocated for data. With a further increase in the number of states, information will need to be recorded in a larger, but low-speed memory, using which the final rate of generation of the secret quantum key will be lower. For this reason, there is a limitation on the amount of recorded information about states in memory caused by the use of high-speed memory.

The marginal losses in the quantum channel in CV-QKD in the presence of collective attacks are 10.2~dB, in the presence of coherent --- 7.5~dB.

\begin{longtable}[c]{|c|c|c|c|c|}
\caption{Parameters used to evaluate the performance of the CV-QKD system. Security parameters are selected in accordance with the work~\cite{Hosseinidehaj2019,Hosseinidehaj2020,Pirandola2021,Pirandola2021a}.}
\label{tab:parameters-2}\\
\hline
\begin{tabular}[c]{@{}c@{}}Parameter\end{tabular} & Description                                                                                                            & \begin{tabular}[c]{@{}c@{}}Value\\ (collective\\attacks)\end{tabular} & \begin{tabular}[c]{@{}c@{}}Value\\ (coherent\\attacks)\end{tabular} & \begin{tabular}[c]{@{}c@{}}Units\end{tabular} \\ \hline
\endfirsthead
\multicolumn{5}{@{}l}%
{{Continuation of Table \thetable}} \\
\hline
\begin{tabular}[c]{@{}c@{}}Parameter\end{tabular} & Description                                                                                                            & \begin{tabular}[c]{@{}c@{}}Value\\ (collective\\attacks)\end{tabular} & \begin{tabular}[c]{@{}c@{}}Value\\ (coherent\\attacks)\end{tabular} & \begin{tabular}[c]{@{}c@{}}Units\end{tabular} \\ \hline
\endhead
$n_{\mathrm{states}}$                                            & number of states                                                                                                  & $6\cdot10^8$                                                            & $6\cdot10^8$                                                           & --                                                          \\ \hline
$\beta$                                                          & \begin{tabular}[c]{@{}c@{}}reconciliation\\efficiency\end{tabular}                                      & 0,95                                                                    & 0,95                                                                   & a.u.                                                    \\ \hline
$\mathrm{FER}$                                                   & \begin{tabular}[c]{@{}c@{}}frame error rate\end{tabular}                        & 0,03                                                                    & 0,03                                                                   & a.u.                                                    \\ \hline
$d$                                                              & \begin{tabular}[c]{@{}c@{}}size of the\\effective\\alphabet\end{tabular}                   & $10^4$                                                                  & $10^4$                                                                 & bit                                                         \\ \hline
$n_{\mathrm{pe}}$                                                & \begin{tabular}[c]{@{}c@{}}number of signals\\ for parameter\\estimation\end{tabular}                      & $6\cdot10^7$                                                            & $6\cdot10^7$                                                           & --                                                          \\ \hline
$f_{\mathrm{et}}$                                                & \begin{tabular}[c]{@{}c@{}}fraction os states\\ for energy tests\end{tabular}                         & 0,2                                                                     & 0                                                                      & a.u.                                                    \\ \hline
$w$                                                              & \begin{tabular}[c]{@{}c@{}}confidence\end{tabular}                                               & 6,34                                                                    & 14,07                                                                  & a.u.                                                    \\ \hline
$\varepsilon_{\mathrm{s}}$                                       & \begin{tabular}[c]{@{}c@{}}smoothness\\ parameter\end{tabular}                                         & $10^{-10}$                                                              & $10^{-43}$                                                             & a.u.                                                    \\ \hline
$\varepsilon_{\mathrm{h}}$                                       & \begin{tabular}[c]{@{}c@{}}parameter that determines\\ hash code match\end{tabular} & $10^{-10}$                                                              & $10^{-43}$                                                             & a.u.                                                    \\ \hline
$\varepsilon$                                                    & \begin{tabular}[c]{@{}c@{}}general security\\parameter\end{tabular}                                                  & $5,6\cdot10^{-9}$                                                       & $1,3\cdot10^{-9}$                                                      & a.u.                                                    \\ \hline
\end{longtable}


Ensuring the security of the CV-QKD protocol with Gaussian modulation against coherent attacks, in turn, requires not only a significant limitation on security criteria, but also an increase in the number of messages to maintain the proper performance level. 

According to the collective attack security criteria set in p.~\ref{sec:col-att} (see also table~\ref{tab:parameters-2}), the number of states is $6 \cdot 10 ^{8}$. Each quantum signal pulse is followed by a reference pulse in such a way that the total number of quantum signal pulses and reference pulses is $6 \cdot 10 ^{8}$ for each block. For each quantum message, the random number generator generates 32 bits: 16 bits each to determine the value of each of the quadratures.

Alice in the process of generating the message writes the package number without taking into account the reference pulses, using 27 bits for this (5 bits are laid down for redundancy to align the word to 4 bytes), as well as a 4-byte number obtained using a software random number generator implemented on the basis of the FPGA Alice module. Bob detects both quadratures of each message received from the channel. Given that half of the states are reference pulses and two quadrature values are recorded for each state, legitimate users within each block write down information about $6 \cdot 10 ^{8}$ quantum state quadrature values. During detection, Bob writes the number of the message (which accounts for 4 bytes) and 4 bytes of information about the two registered values of quadratures of quantum messages to high-speed DDR memory, as well as 4 bytes of information about the two registered values of the quadratures of the reference pulses.

Information about the registered reference pulse quadratures is used to compensate for the phase shift of quantum messages as a result of transmission over a quantum channel, after which this information is deleted from memory. Every ten states, Bob randomly selects one to be used for channel characterization (for the parameter estimation procedure). For the remaining messages, insignificant bits are discarded in the part containing information about quadratures, as a result of which eight bits remain out of 32 bits of information. On average, each package has 5.3 bytes of information. The time taken for detection and storage is 12 seconds.

To evaluate the channel, states are disclosed for which insignificant bits have not been discarded. After that, errors are corrected using multi-level encoding and multi-stage decoding~\cite{VanAssche2004,Mani2021,Wen2021}. During error correction, three of the four bits are revealed, thereby reducing the bit sequence to a length of $5.4\cdot 10^{8}$ bits.

After the error correction is completed, universal hashing is used to exhaustively verify that all Alice and Bob sequences are the same (confirmation procedure). Hashing of the key with corrected errors is performed both in Alice's and Bob's blocks. As a result of hashing, users still have hash codes on their hands. Bob sends the received hash code to Alice. She then compares the values of two hash codes: the one calculated in her block and the one received from Bob. The result of the comparison is then transmitted to him. If the values do not match, then the processing of this sifted key stops without generating a secret key. The key that has not passed the confirmation procedure is erased from memory on both sides.

A 2-universal hash function is used at the privacy amplification step. The input is a key with corrected errors. The corrected key is loaded in blocks. The ratio of the length of the output sequence after privacy amplification procedure to the length of the input sequence is 1:66.

Since the post-processing of the sequence is faster than writing information to memory during detection, these processes can be performed in parallel. 

\section{Conclusion}
\label{sec:conclusion}
In this paper, we have carried out a theoretical analysis of the performance of a realistic CV-QKD system. Our estimates show that performance can be maintained with losses as low as 10~dB in the most common assumption of collective attacks. In the presence of coherent attacks, there is a noticeable drop in allowable losses (down to 7~dB), and at the same time, tougher security criteria must be taken into account, which are still quite difficult to satisfy in practice. Further work will be focused on creating an experimental setup in accordance with what is described in the article and evaluating the performance of a real system.

\section*{Acknowledgements}
The work was done by Leading Research Center "National Center for Quantum Internet" of ITMO University by order of JSCo Russian Railways.


\end{document}